\documentclass[11pt]{article}
\usepackage[dvipsnames]{xcolor}
\usepackage{times}
\usepackage{amsmath,amsthm,amssymb,setspace,enumitem,epsfig,titlesec,verbatim,array,eurosym,multirow}
\usepackage[sort&compress,numbers]{natbib}
\usepackage[footnotesize,bf]{caption}
\usepackage[margin=2.5cm, includefoot, footskip=30pt]{geometry}
\usepackage{standalone}
\usepackage{tikz}
\usepackage{subcaption}
\usepackage{hyperref}
\usepackage{tabularx}
\usepackage{booktabs}
\usepackage{blkarray}
\usepackage[ruled,vlined]{algorithm2e}
\smallskip 

\usepackage{tikz}
\usetikzlibrary{arrows}
\usepackage{minitoc}
\usepackage{graphicx}
\usepackage{hyperref}
\usepackage{subcaption}
\usepackage{multirow}
\usepackage{multicol}
\usepackage{standalone}
\usepackage{pdfpages}

\newcommand{\figref}[1]{{\textbf{Fig.~\ref{#1}}}}

\def\tft{\texttt{TFT}}
\def\tftt{\texttt{TF2T}}
\def\ttft{\texttt{2TFT}}

\def\alld{\texttt{ALLD}}

\def\SI{\textbf{Supporting Information}}
\def\methods{\textbf{Material and Methods}}

\titleformat{\section}{\sffamily \fontsize{12}{15}\bfseries}{\thesection}{0.4em}{}
\titleformat{\subsection}{\sffamily\fontsize{11}{15}\bfseries}{\thesubsection}{0.4em}{}

\title{\bfseries \sffamily \Large Conditional cooperation with longer memory}

\author{Nikoleta E. Glynatsi$^{1,*}$, Martin A. Nowak$^2$, Christian Hilbe$^1$\\[0.3cm]
$^{\rm 1}$Max Planck Research Group on the Dynamics of Social Behavior,\\ Max Planck Institute for Evolutionary Biology, Pl\"{o}n, Germany \\
$^{\rm 2}$Department of Mathematics,
Department of Organismic and Evolutionary Biology,\\ Harvard University, Cambridge, USA\\
$^*$To whom correspondence should be addressed. E-mail: glynatsi@evolbio.mpg.de
}
\date{}

\begin{document}

\maketitle

~\\[0.5cm]
\noindent
{\bf 
Direct reciprocity is a wide-spread mechanism for evolution of cooperation. In
repeated interactions, players can condition their behavior on previous
outcomes. A well known approach is given by reactive strategies, which respond
to the co-player's previous move. Here we extend reactive strategies to longer
memories. A reactive-$n$ strategy takes into account the sequence of the last $n$
moves of the co-player. A reactive-$n$ counting strategy records how often the
co-player has cooperated during the last $n$ rounds. We derive an algorithm to
identify all partner strategies among reactive-$n$ strategies. We give explicit
conditions for all partner strategies among reactive-2, reactive-3 strategies,
and reactive-$n$ counting strategies. Partner strategies are those that ensure
mutual cooperation without exploitation. We perform evolutionary simulations and
find that longer memory increases the average cooperation rate for reactive-$n$
strategies but not for reactive counting strategies. Paying attention to the
sequence of moves is necessary for reaping the advantages of longer memory.
}\\[1cm]

\noindent
{\it Keywords:} Evolutionary game theory, direct reciprocity, evolution of cooperation, prisoner's dilemma

\clearpage
\newpage


\noindent
{\bf Significance statement.}
In repeated interactions, people tend to cooperate conditionally. They are
influenced by whether others cooperate with them, and react accordingly. Direct
reciprocity is based on repeated interactions between two players. Nice
strategies are those that are never the first to defect. Consequently, they
never seek to exploit the other. Partner strategies are nice strategies which
can sustain full cooperation as a Nash equilibrium. If you interact with such a
partner then you maximize your own payoff by full cooperation. Therefore,
partners resolve social dilemmas. Here we characterize all nice and all partner
strategies among longer memory reactive strategies. Our results show that
natural selection chooses partners. It pays to be nice.


\section*{Introduction}

\begin{figure}[t]
  \centering
  \includegraphics[width=0.75\textwidth]{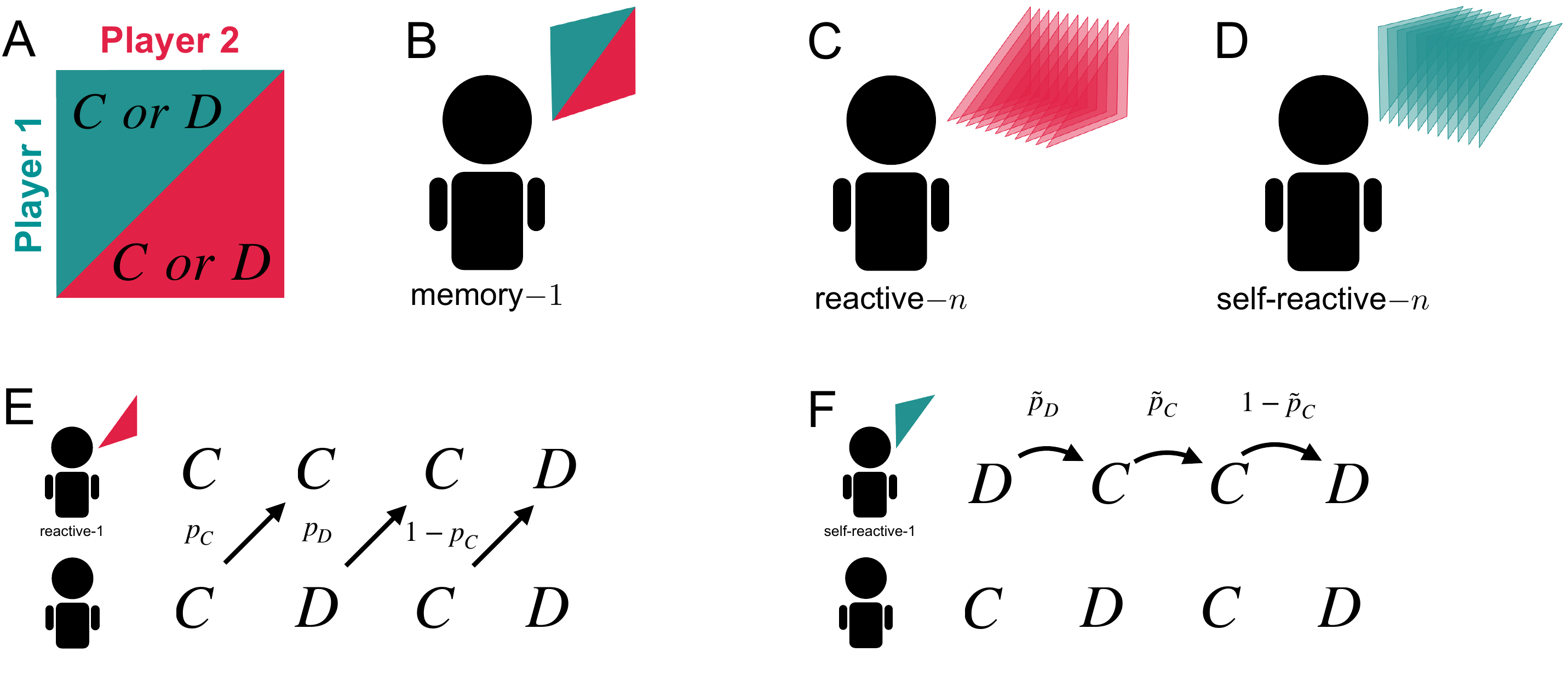}
  \caption{\textbf{The repeated prisoner's dilemma among players with finite memory.}
  \textbf{A,} In the repeated prisoner's dilemma, in each round two players independently decide whether to cooperate~($C$) or to defect~($D$). 
   \textbf{B,} When players adopt memory-1 strategies, their decisions depend on the entire outcome of the previous round. That is, they consider both their own and the co-player's previous action. 
   \textbf{C,} When players adopt a reactive-$n$ strategy, they make their decisions based on the co-player's actions during the past $n$ rounds. 
   \textbf{D,} A self-reactive-$n$ strategy is contingent on the player's own actions during the past $n$ rounds. 
   \textbf{E,} To illustrate these concepts, we show a game between a player with a reactive-$1$ strategy (top) and an arbitrary player (bottom). 
   Reactive-1 strategies can be represented as a vector  $\mathbf{p} \!=\! (p_C, p_D)$. 
   The entry $p_C$ is the probability of cooperating given the co-player cooperated in the previous round.
   The entry $p_D$ is the cooperation probability after the co-player defected. 
   \textbf{E,} Now, the top player adopts a self-reactive-1 strategy, $\mathbf{\tilde p}\!=\!(\tilde p_C, \tilde p_D)$. 
   Here, the bottom player's cooperation probabilities depend on their own previous action.
   }\label{fig:conceptual_figure_model}
\end{figure}

To a considerable extent, human cooperative behavior is governed by direct reciprocity~\cite{melis:ptrs:2010,rand:TCS:2013}. 
This mechanism for cooperation can explain why people return favors~\citep{neilson:JEBO:1999}, why they show more effort in group tasks when others do~\citep{fischbacher:AER:2010}, or why they stop cooperating when they feel exploited~\citep{hilbe:ncomms:2014,Xu:NComms:2016}. 
The main theoretical framework to describe reciprocity is the repeated prisoner's dilemma~\cite{axelrod:AAAS:1981, nowak:Science:2006, sigmund2010,Garcia:FRAI:2018,hilbe:Nature:2018,Rossetti:ETH:2023}. 
This game considers two individuals, referred to as players, who repeatedly decide whether to cooperate or to defect with one another~(\figref{fig:conceptual_figure_model}\textbf{A}). 
Both players prefer mutual cooperation to mutual defection. 
Yet given the co-player's action, each player has an incentive to defect. 
One common implementation of the prisoner's dilemma is the donation game. 
Here, cooperation simply means to pay a cost $c\!>\!0$ for the co-player to get a benefit $b\!>\!c$. 
Despite the simplicity of these games, they can give rise to remarkable dynamical patterns. 
These patterns have been explored in numerous studies~\citep{frean:PRSB:1994,killingback:PRSB:1999,hauert:JTB:2002b,kurokawa:PRSB:2009,pinheiro:PLoSCB:2014,garcia:jet:2016,mcavoy:PRSA:2019,Kraines:TaD:1989,nowak:Nature:1993,imhof:PNAS:2005,grujic:jtb:2012,van-segbroeck:prl:2012,press:PNAS:2012,stewart:pnas:2013,Toupo:IJBC:2014,stewart:pnas:2014, akin:EGADS:2016,glynatsi:scientific:2020,chen:PNASnexus:2023}.
Some of this literature describes how the evolution of cooperation depends on the game parameters, such as the benefit of cooperation, or the frequency with which errors occur~\citep{boyd:JTB:1989,Hao:PRE:2015,Zhang:GEB:2018,Mamiya:PRE:2020}. 
Others describe the effect of different learning dynamics~\citep{stewart:games:2015,Mcavoy:PNAS:2022}, of population structure~\citep{brauchli:JTB:1999,szabo:PRE:2000b,allen:AmNat:2013,szolnoki:scirep:2014}, or of the strategies that players are permitted to use~\citep{baek:scientific:2016}.  

Strategies of the repeated prisoner's dilemma can vary in their complexity.
While some are straightforward to implement, like always defect, many others are more sophisticated~\cite{harper:PLOSONE:2017, knight:PLOSONE:2018}.
To quantify a strategy's complexity, it is common to resort to the number of past rounds that the player needs to remember. 
Unconditional strategies like `always defect' or `always cooperate' are said to be \mbox{memory-0}. 
Strategies that only depend on the previous round, such as `Tit-for-Tat'~~\cite{axelrod:AAAS:1981,Duersch:IJGT:2013} or `Win-Stay Lose-Shift'~\cite{Kraines:TaD:1989,nowak:Nature:1993}, are \mbox{memory-1}~~(\figref{fig:conceptual_figure_model}\textbf{B}). 
Similarly, one can distinguish strategies that require more than one round of memory, or strategies that cannot be implemented with finite memory~\cite{Garcia:FRAI:2018}. 

Traditionally, most theoretical research on the evolution of reciprocity focuses on memory-1 strategies~\citep{nowak:Nature:1993,imhof:PNAS:2005,grujic:jtb:2012,van-segbroeck:prl:2012,press:PNAS:2012,stewart:pnas:2013,Toupo:IJBC:2014,stewart:pnas:2014, akin:EGADS:2016,glynatsi:scientific:2020,chen:PNASnexus:2023}. 
Although one-round memory can explain some of the empirical regularities in human behavior~\cite{engle:ET:2006, dal:AER:2011, camera:GEB:2012, bruttel:TD:2012,Montero-Porras:SciRep:2022},  people often take into account more than the last round~\cite{romero:EER:2018}.
Longer memory seems particularly relevant for noisy games, where people occasionally defect because of unintended errors~\cite{fudenberg:AER:2012}. 
However, a formal analysis of strategies with more than one-round memory has been difficult for two reasons. 
First, as the memory length $n$ increases, strategies become harder to interpret. 
For example, because two consecutive rounds of the prisoner's dilemma allow for 16 possible outcomes, memory-2 strategies need to specify 16 conditional cooperation probabilities~\citep{hauert:PRSB:1997}. 
Although some of the resulting strategies have an intuitive interpretation, such as `Tit-for-Two-Tat'~\citep{axelrod:AAAS:1981}, many others are difficult to make sense of. 
Second, the number of strategies, and the time it takes to compute their payoffs, increases dramatically in $n$. 
For example, for memory-1, there are $2^4\!=\!16$ deterministic strategies (strategies that do not randomize between different actions). 
When both players adopt memory-1 strategies, computing their payoffs requires the inversion of a $4\!\times\!4$ matrix~\cite{sigmund2010}. 
After increasing the memory length to memory-2, there are $2^{16}\!=\!64,536$ deterministic strategies, and payoffs now require the inverse of a $16\!\times\!16$ matrix. 
Probably for these reasons, previous studies considered simulations for small $n$~\citep{hauert:PRSB:1997,stewart:scientific:2016,Murase:PLoSCompBio:2023a}, or they analyzed the properties of a few selected higher-memory strategies~\citep{hilbe:PNAS:2017,ueda:RSOP:2021,li:NatureCompSci:2022}. 

To make progress, we focus on an easy-to-interpret subset of memory-$n$ strategies, the {\it reactive}-$n$ strategies. 
Capturing the basic premise of conditional cooperation, they only depend on the {\it co-player's} actions during the last $n$ rounds~(\figref{fig:conceptual_figure_model}\textbf{C,E}). 
While it has been difficult to explicitly characterize all Nash equilibria among the memory-$n$ strategies, we show that such a characterization is possible for reactive-$n$ strategies. 
Our results rely on a central insight, motivated by previous work by Press~\&~Dyson~\citep{press:PNAS:2012}: 
if one player adopts a reactive-$n$ strategy, the other player can always find a best response among the deterministic {\it self-reactive}-$n$ strategies. 
Self-reactive-$n$ strategies are remarkably simple. 
They only depend on the player's own previous $n$ moves (\figref{fig:conceptual_figure_model}\textbf{D,F}).
Based on this insight, we study all reactive-$n$ strategies that sustain full cooperation in a Nash equilibrium (the so-called {\it partner strategies}). 
We provide a full characterization for $n\!=\!2$ and $n\!=\!3$.
Even stronger results are feasible when we restrict attention to so-called {\it counting strategies}.  
Such strategies only react to how often the co-player has cooperated in the last $n$ rounds (irrespective of the exact timing of cooperation). 
For the donation game, we characterize the partners among the counting strategies for arbitrary~$n$. 
The resulting conditions are straightforward to interpret:
For every defection of the co-player in memory, the focal player's cooperation rate needs to drop by $c/(nb)$.
To further assess the relevance of partner strategies for the evolution of cooperation, we conduct extensive simulations for $n\!\in\!\{1,2,3\}$. 
Our findings indicate that the evolutionary process strongly favors partner strategies, and that these strategies are crucial for cooperation. 

Overall, our results provide important insights into the logic of conditional cooperation when players have more than one-round memory. 
We show that partner strategies exist for all repeated prisoner's dilemmas and for all memory lengths. 
To be stable, however, these strategies need to be sufficiently responsive to the co-player's previous actions.


\section*{Results}


\textbf{Model and notation.}
We consider a repeated game between two players, player~1 and player~2.
Each round, players can choose to cooperate ($C$) or to defect ($D$). 
If both players cooperate, they receive the reward~$R$, which exceeds the (punishment) payoff~$P$ for mutual defection. 
If only one player defects, the defector receives the temptation payoff~$T$, whereas the cooperator ends up with the sucker's payoff~$S$. 
We assume payoffs satisfy the typical relationships of a prisoner's dilemma, $T \!>\! R \!>\! P \!>\! S$ and $2 R \!>\! T \!+\! S$. 
Therefore, in each round, mutual cooperation is the best outcome for the pair, but players have some incentive to defect. 
The players' aim is to maximize their average payoff per round, across infinitely many rounds.
To make results easier to interpret, it is sometimes instructive to look at a particular variant of the prisoner's dilemma, the donation game. 
Here, cooperation means to pay a cost $c\!>\!0$ for the co-player to get a benefit $b\!>\!c$.
The resulting payoffs are \(R \!\!=\! b\! -\! c, S \!=\! -c, T \!=\! b, P\!
=\! 0\). 
To illustrate our results, we focus on the donation game in the following. 
However, most of our findings are straightforward to extend to the general prisoner's dilemma (or to other repeated $2\!\times\!2$ games, see \SI). 

We consider players who use strategies with finite memory. 
To describe such strategies formally, we introduce some notation. 
The last $n$ actions of each player  $i\! \in\! \{1, 2\}$ are referred to as the player's {\it $n$-history}. 
We write this $n$-history as a tuple $\mathbf{h}^i\!=\!(a^i_{-n},\ldots,a^i_{-1})\!\in\!\{C,D\}^n$. 
Each entry $a^i_{-k}$ corresponds to player $i$'s action $k$ rounds ago. 
We use $H^i$ for the set of all such $n$-histories. 
This set contains $|H^i|\!=\!2^{n}$ elements. 
Based on this notation, we can define a {\it reactive}-$n$ {\it strategy} for player 1 as a vector $\mathbf{p}\!=\!(p_\mathbf{h})_{\mathbf{h}\in H^2} \!\in\! [0, 1]^{2^n}$. 
The entries $p_\mathbf{h}$ correspond to player 1's cooperation probability in any given round, contingent on player 2's actions during the last $n$ rounds. 
The strategy is called pure or deterministic if any entry is either zero or one. 
We note that the above definition leaves player 1's moves during the first $n$ rounds unspecified. 
However, in infinitely repeated games without discounting, these initial moves tend to be inconsequential. 
Hence, we neglect them in the following.

For \(n\!=\!1\), the above definition recovers the classical format of reactive-1 strategies~\cite{sigmund2010}, \(\mathbf{p}\!=\!(p_C, p_D)\). 
Here, $p_C$ and $p_D$ are the player's cooperation probability given that the co-player cooperated or defected in the previous round, respectively. 
This set contains, for example, the strategies of unconditional defection, \alld~$=\!(0,0)$, and Tit-for-Tat, \tft~$=\!(1,0)$. 
The next complexity class is the set of reactive-2 strategies, $\mathbf{p}\!=\!(p_{CC},p_{CD},p_{DC},p_{DD})$.
In addition to \alld{} and \tft{}, this set contains, for instance, the strategies Tit-for-Two-Tat, \tftt~$=\!(1,1,1,0)$ and Two-Tit-for-Tat,~\ttft$=\!(1,0,0,0)$. 
Similar examples exist for $n\!>\!2$. 
When both players adopt reactive-$n$ strategies (or more generally, memory-$n$ strategies), it is straightforward to compute their expected payoffs, by representing the game as a Markov chain. 
The respective procedure is described in the \SI{}.  

Herein, we are particularly interested in those reactive-$n$ strategies that sustain full cooperation. 
Such strategies ought to have two properties. 
First, they ought to be {\it nice}, meaning that they are never the first to defect~\citep{axelrod:AAAS:1981}.
This property ensures that two players with nice strategies fully cooperate. 
In particular, if $\mathbf{h}_C$ is a co-player's $n$-history that consists of $n$ bits of cooperation, a nice strategy needs to respond by cooperating with certainty, $p_{\mathbf{h}_C}\!=\!1$.  
Second, the strategy ought to form a {\it Nash equilibrium}, such that no co-player has an incentive to deviate. 
Strategies that have both properties are called {\it partner strategies}~\citep{Hilbe:GEB:2015} or {\it partners}.
The partners among the reactive-1 strategies are well known. 
For the donation game, partners are those strategies with $p_C\!=\!1$ and $p_D\!\le\!1\!-\!c/b$~\citep{akin:EGADS:2016}. 
However, a general theory of partners for $n\!\ge\!2$ is lacking. 
This is what we aim to derive in the following. 
In the main text, we provide the main intuition for our results; all proofs are in the \SI.\\

 
\noindent
\textbf{An algorithm to identify partners among the reactive-$n$ strategies.} 
It is comparably easy to verify whether a reactive-$n$ strategy $\mathbf{p}$ is nice. 
Demonstrating that the strategy is also a Nash equilibrium, however, is far less trivial. 
In principle, this requires uncountably many payoff comparisons. 
We would have to show that if player 2's strategy is fixed to $\mathbf{p}$, no other strategy $\sigma$ for player~1 can result in a higher payoff. That is, player 1's payoff needs to satisfy $\pi^1(\sigma,\mathbf{p})\!\le\!\pi^1(\mathbf{p},\mathbf{p})$ for all $\sigma$. 
Fortunately, this task can be simplified considerably. 
Already Press~\& Dyson~\cite{press:PNAS:2012} showed that it is sufficient to test only those $\sigma$ with at most $n$ rounds of memory. 
Based on two insights, we can even further restrict the search space of strategies $\sigma$ that need to be tested.

\begin{figure}[t]
  \centering
  \includegraphics[width=\textwidth]{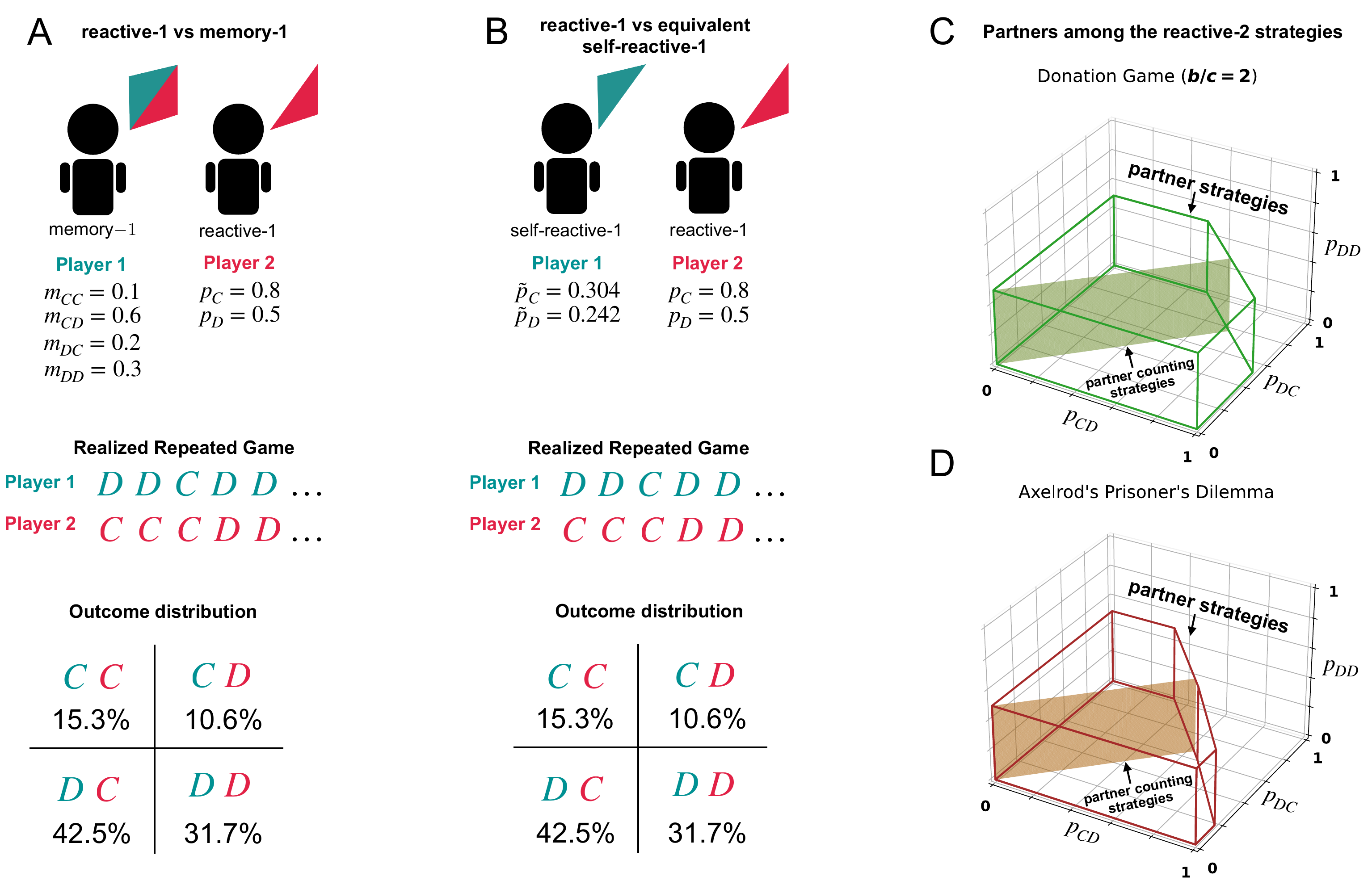}
  \caption{
  \textbf{Characterizing the partners among the reactive-$n$ strategies.} 
{\bf A,B,} To characterize the reactive-$n$ partner strategies, we prove the following result. 
Suppose the focal player adopts a reactive-$n$ strategy. 
Then, for any strategy of the opponent (with arbitrary memory), one can find an associated self-reactive-$n$ strategy that yields the same payoffs. 
Here, we show an example where player~1 uses a reactive-1 strategy against player~2 with a memory-1 strategy. 
Our result implies that can switch to a well-defined self-reactive-1 strategy. 
This switch leaves the outcome distribution unchanged.
In both cases, players are equally likely to experience mutual cooperation, unilateral cooperation, or mutual defection in the long run. 
\textbf{C,} Based on this insight, we can explicitly characterize the reactive-2 partner strategies (with $p_{CC}\!=\!1$). 
Here, we represent the corresponding conditions~\eqref{eq:two_bit_conditions} for a donation game with $b/c\!=\!2$. 
Among the reactive-2 strategies, the counting strategies correspond to the subset with $p_{CD}\!=\!p_{DC}$. 
Counting strategies only depend on how often the co-player cooperated in the past, not on the timing of cooperation.
\textbf{D,} Similarly, we can also characterize the reactive-2 partner strategies for the general prisoner's dilemma. 
Here, we use the values of Axelrod~\citep{axelrod:AAAS:1981}.
}\label{fig:conceptual_figure_results}
\end{figure}

First, suppose player~1 uses some arbitrary strategy $\sigma$ against player~2 with reactive-$n$ strategy \mbox{$\mathbf{p}\!=\!(p_\mathbf{h})_{\mathbf{h}\in H^1}$}. 
Then we prove that instead of $\sigma$, player~1 may switch to a {\it self-reactive}-$n$ strategy $\mathbf{\tilde{p}}$ without changing either player's payoffs. 
When adopting a self-reactive strategy, player~1 only takes into account her own actions during the last $n$ rounds, 
$\mathbf{\tilde{p}} \!=\! (\tilde{p}_\mathbf{h})_{\mathbf{h} \in H^1}$.
In particular, if $\sigma$ is a best response to $\mathbf{p}$, then there is an associated self-reactive strategy $\mathbf{\tilde p}$ that is also a best response. 
This result follows the same intuition as a similar result of Press \& Dyson~\cite{press:PNAS:2012}: 
if there is a part of the joint history that player~2 does not take into account, player~1 gains nothing by considering that part of the history. 
In our case, because player~2 only considers the last $n$ actions of player~1, it is sufficient for player~1 to do the same.
\figref{fig:conceptual_figure_results}\textbf{A,B} provides an illustration.  
There, we depict a game in which player~1 adopts a memory-1 strategy against a reactive-1 opponent. 
Due to the above result, we can find an equivalent self-reactive-1 strategy for player~1. 
While that self-reactive strategy is simpler, on average it induces the same game dynamics. 
Hence, it results in identical payoffs. 

The above result guarantees that for any reactive-$n$ strategy, there is always a best response among the self-reactive-$n$ strategies. 
In a second step, we prove that such a best response can always be found among the {\it deterministic} self-reactive-$n$ strategies. 
This reduces the search space for potential best responses further, from an uncountable set to a finite set of size $2^{2^n}$. 
For $n\!=\!2$, this leaves us with 16 self-reactive strategies to test. 
For $n\!=\!3$, we end up with (at most) 256 strategies. 
While this may still appear to be a large number, many of the different strategies impose redundant constraints on partner strategies.
This redundancy further reduces the number of conditions a partner needs to satisfy.\\

\begin{figure}[t]
  \centering
  \includegraphics[width=\textwidth]{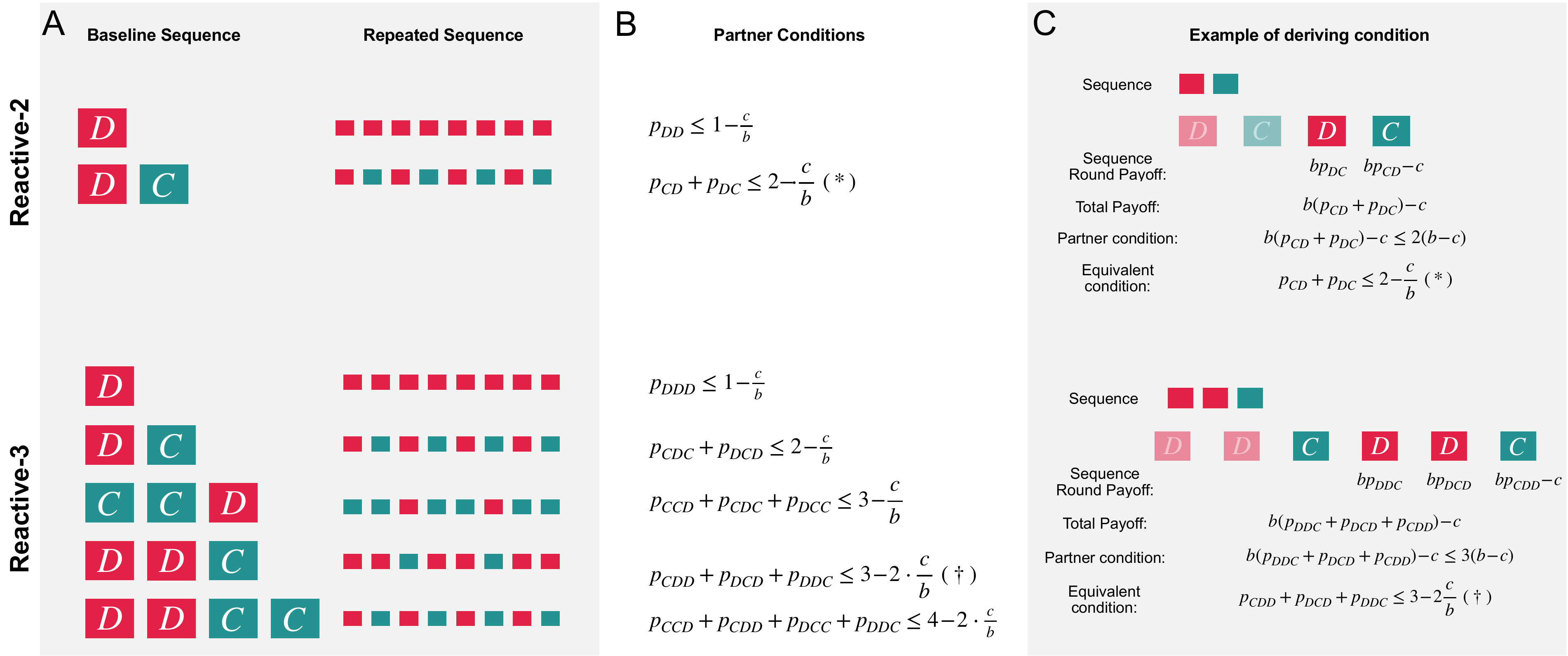}
  \caption{
  \textbf{Conditions for partners among reactive-$2$ and reactive-$3$ strategies.}
  {\bf A, } Within the set of pure self-reactive strategies, there are certain
  strategies where the way they behave can be described as playing a unique
  sequence. Since their action does not depend on the co-player's history, the
  strategy plays this sequence indefinitely. For example, in the case of \(n=2\),
  the pure self-reactive strategy \(\mathbf{\tilde{p}}=(0, 1)\), which alternates,
  can be described as playing the sequence \((D, C)\).
  {\bf B,} We have proven that to characterize reactive-\(n\) partner strategies, one only
  needs to check deviations towards self-reactive-\(n\) strategies. Thus, for a
  nice reactive strategy \(p\) to be a partner, it is necessary that none of these
  sequence-playing self-reactive strategies can achieve a higher payoff against
  \(p\) than \(p\) does against itself. The conditions of partner strategies, for
  \(n=2\), and \(n=3\), respectively for each of the sequences in panel {\bf
  A}, are shown in panel {\bf B}. These conditions are necessary, but
  furthermore, we have shown that these are also sufficient conditions (see
  \SI).
  {\bf C,} To derive the conditions, we need to consider the payoff that a sequence player
  achieves against a reactive strategy. In the top panel of panel {\bf C}, we
  illustrate an example for \(n=2\), against \(\mathbf{p} = (1, p_{CD}, p_{DC},
  p_{DD})\) and the sequence \((D, C)\). In the third turn, the sequence player
  receives a benefit \(b\) with a probability of \(p_{DC}\), and no cost since the
  sequence player did not cooperate. In the fourth turn, the player receives
  \(p_{DC} \cdot b - c\), and thereafter these two payoffs are repeated forever.
  Thus, the total payoff of the sequence player with memory two is given by what
  they receive every two turns, which is \(p_{DC} \cdot b - c\). This payoff needs
  to be smaller or equal than what a partner strategy achieves against another
  nice strategy, which is \(2(b\!-\!c)\).
  In the bottom panel of panel {\bf C}, we illustrate an example for $n=3$.
  }\label{fig:conceptual_figure_conditions}
\end{figure}

\noindent
\textbf{Partners among the reactive-2 and the reactive-3 strategies.}
To illustrate the above algorithm, we first characterize the partners among the reactive-$2$ strategies. 
To this end, we note that it is straightforward to compute the payoff of a specific self-reactive-2 strategy against a general reactive-2 strategy $\mathbf{p}$ (see \SI{} for details). 
By computing the payoffs of all 16 pure self-deterministic strategies $\mathbf{\tilde p}$, and by requiring $\pi^1(\mathbf{\tilde p},\mathbf{p}) \!\le\! \pi^1(\mathbf{p},\mathbf{p})$ for all of them, we end up with only three conditions. Specifically, we prove that $\mathbf{p}$ is a partner if and only if
\begin{equation}\label{eq:two_bit_conditions}
  p_{CC} = 1, \qquad  \frac{p_{CD} + p_{DC}}{2} \le 1 - \frac{1}{2} \!\cdot\! \frac{c}{b}, \qquad  p_{DD} \leq 1\!-\! \frac{c}{b}.
\end{equation}
The above conditions define a three-dimensional polyhedron within the space of all nice reactive-2 strategies~(\figref{fig:conceptual_figure_results}\textbf{C}).
The condition $p_{CC}\!=\!1$ follows from the requirement that the strategy ought to be nice. 
As long as the co-player cooperates, the reactive-$n$ player goes along. 
The other two conditions imply that for each defection in memory, the player's cooperation rate decreases by $c/(2b)$.
Interestingly, in cases with a mixed $2$-history (one cooperation, one defection), the above conditions suggest that the exact timing of cooperation does not matter. 
It is only required that the two cooperation probabilities $p_{CD}$ and $p_{DC}$ are sufficiently small {\it on average}. 
Notably, the above conditions also imply that to check whether a given reactive-2 strategy is a partner, it suffices to check two deviations. 
These deviations are the strategy that strictly alternates between cooperation and defection (yielding the first inequality), and \alld~(yielding the second inequality) (\figref{fig:conceptual_figure_conditions}). 
We note that this last implication is specific to the donation game. 
For the general prisoner's dilemma (depicted in \figref{fig:conceptual_figure_results}\textbf{D}), there are more than two inequalities that need to be satisfied (see~\SI{}).

Analogously, we can also characterize the partners among the reactive-3 strategies. 
A reactive-3 strategy is defined by the vector 
$
\mathbf{p} = (p_{CCC},\, p_{CCD},\, p_{CDC},\, p_{CDD},\, p_{DCC},\, p_{DCD},\, p_{DDC},\, p_{DDD}).
$
It is a partner strategy if and only if
\begin{align}\label{eq:three_bit_conditions}
  \begin{split}
  p_{CCC} & = 1 \\[0.2cm]
  \frac{p_{CDC} + p_{DCD}}{2} & \leq 1 - \frac{1}{2} \cdot \frac{c}{b} \\[0.2cm]
  \frac{p_{CCD} + p_{CDC} + p_{DCC}}{3} & \leq 1 - \frac{1}{3} \cdot \frac{c}{b} \\[0.2cm]
  \frac{p_{CDD} + p_{DCD} + p_{DDC}}{3} & \leq 1 - \frac{2}{3} \cdot \frac{c}{b} \\[0.2cm]
  \frac{p_{CCD} + p_{CDD} + p_{DCC} + p_{DDC}}{4}  & \leq 1 - \frac{1}{2} \cdot \frac{c}{b}  \\[0.2cm]
  p_{DDD} & \leq 1\!-\! \frac{c}{b}
  \end{split}
\end{align}
These conditions follow a similar logic as in the previous case with $n\!=\!2$:
for every co-player's defection in memory, the respective cooperation
probability needs to be diminished proportionally. These conditions
conditions also imply that to check whether a given reactive-3 strategy is a
partner, it suffices to check five deviations. Similarly to the previous case,
two of these deviations include the strategy that strictly alternates between
cooperation and defection, and \alld. The rest of the conditions arise from
deviations towards sequence-playing self-reactive strategies, where the
sequences are \((CCD)\), \((DCC)\), and \((DDCC)\) (\figref{fig:conceptual_figure_conditions}).
For $n\!=\!3$, there are now more conditions to consider than in the previous
case, and these conditions become even more complex for the general prisoner's
dilemma. Given these complexities, we do not present conditions for reactive-$n$
partner strategies beyond $n\!=\!3$, even though the algorithm presented in the
previous section still applies.\\


\noindent
\textbf{Partners among the reactive-$n$ counting strategies.}
We can more easily generalize these formulas to the case of arbitrary $n$ if we further restrict the strategy space. 
In the following, we consider reactive-$n$ {\it counting strategies}. 
These strategies take into account how often the co-player cooperated during the past $n$ rounds. 
However, they do not consider in which of the past $n$ rounds the co-player cooperated. 
In the following, we represent such strategies as a vector $\mathbf{r}\!=\!(r_i)_{i \in \{n, n -1, \dots, 0\}}$. 
Each entry \(r_i\) indicates the player's cooperation probability if the co-player cooperated \(i\) times during the last \(n\) rounds. 
Note that any reactive-1 strategy $\mathbf{p}\!=\!(p_{C}, p_{D})$ is a counting strategy by definition. 
However, for larger $n$, the set of counting strategies is a strict subset of the reactive-$n$ strategies.
For example, for $n\!=\!2$, counting strategies are those strategies that satisfy $p_{CD}\!=\!p_{DC}\!=:\!r_1$. 
As a result, the partners among the counting strategies form a 2-dimensional plane within the 3-dimensional polyhedron of reactive-2 partner strategies (\figref{fig:conceptual_figure_results}\textbf{C,D}).

For the donation game among players with counting strategies, it is possible to characterize the set of partner strategies for arbitrary $n$. We find that a counting strategy $\mathbf{r}$ is a partner if and only if
\begin{equation} \label{eq:counting}
  r_n = 1 \qquad \text{and} \qquad r_{n - k} \le 1\! -\! \frac{k}{n} \!\cdot\! \frac{c}{b}~~ \text{ for }~k \!\in\! \{1, 2, \dots, n\}.
\end{equation}
That is, for every defection of the opponent in memory, the maximum cooperation probability needs to be reduced by $c/(nb)$.
It is worth to highlight that this result is general. 
These strategies are Nash equilibria even if players are allowed to deviate towards strategies that do not merely count the co-player's cooperative acts, or towards strategies that take into account more than the last $n$ rounds.\\


\noindent
\textbf{Evolutionary Dynamics.}
With our previous equilibrium analysis we have identified the strategies that can sustain cooperation in principle. 
In a next step, we determine whether these strategies can evolve in the first place. 
Here, we no longer presume that individuals would play equilibrium strategies. 
Rather they initially implement some random behavior. 
Over time, however, they adapt their strategies based on social learning.
To model this learning process, we consider a population of individuals who update their strategies based on pairwise comparisons. 
The efficacy of the resulting learning process is determined by a strength of selection parameter~$\beta$. 
The larger $\beta$, the more likely individuals imitate strategies with a higher payoff. 
In addition, mutations occasionally introduce new strategies.
We describe the exact setup of this learning process in the \methods{} section.
As we explain there, the process is particularly easy to explore when mutations are rare~\cite{fudenberg:JET:2006,wu:JMB:2012,imhof:royal:2010,mcavoy:jet:2015}. 
In that case, the population is typically homogeneous, such that all players adopt the same (resident) strategy.   
Once a new mutant strategy appears, this strategy fixes or goes extinct before the next mutation happens. 
Evolutionary processes with rare mutations can be simulated more efficiently because there is an explicit formula for the mutant's fixation probability~\citep{nowak:Nature:2004}. 

\begin{figure}[t]
  \centering
  \includegraphics[width=\textwidth]{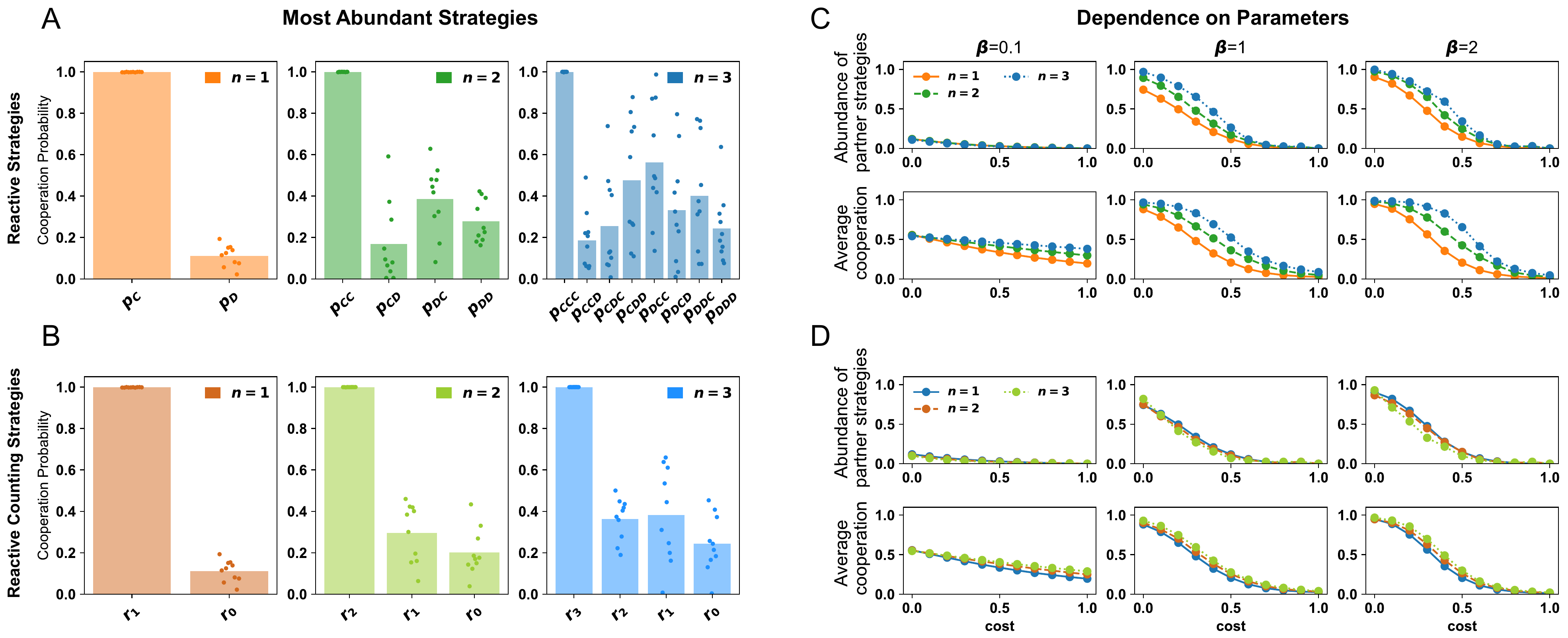}
  \caption{\textbf{Evolutionary dynamics of reactive-$n$ strategies.}
  To explore the evolutionary dynamics among reactive-$n$ strategies, we run simulations based on the
  method of Imhof and Nowak~\cite{imhof:royal:2010}. 
  This method assumes rare mutations. 
  Every time a mutant strategy appears, it goes extinct or fixes before the arrival of the next mutant strategy. 
  {\bf A,B,} We run ten independent simulations for reactive-$n$ strategies and for reactive-$n$ counting strategies. 
  For each simulation, we record the most abundant strategy (the strategy that resisted most mutants). 
  The respective average cooperation probabilities are in line with the conditions for partner strategies. 
  {\bf C,D,} With additional simulations, we explore the average abundance of partner strategies and the population's average cooperation rate. 
For a given resident strategy to be classified as a partner by our simulation, it needs to satisfy all inequalities in the respective definition of partner strategies. 
In addition, it needs to cooperate after full cooperation with a probability of at least 95\%.  
For all considered parameter values, we only observe high cooperation rates when partner strategies evolve. 
Simulations are based on a donation game with \(b\!=\!1\),  \(c\!=\!0.5\), a selection strength $\beta\!=\!1$
and a population size $N\!=\!100$, unless noted otherwise. For $n$ equal to 1 and 2, simulations are run for \(T\!=\! 10 ^ 7\) time steps. For $n\!=\!3$ we use \(T\!=\! 2 \!\cdot\!10 ^ 7\) time steps.
}\label{fig:evolutionary_results}
\end{figure}

The results of these simulations are shown in \figref{fig:evolutionary_results}. 
First, we explore which reactive-$n$ strategies evolve for a fixed set of game parameters. 
Here, we only vary the strategies' memory length $n$, and whether mutations can introduce all reactive-$n$ strategies, or counting strategies only. 
For ten independent simulations, \figref{fig:evolutionary_results}\textbf{A,B} displays the most abundant strategy for each simulation run (those are the strategies that prevent the largest number of mutants from taking over). 
We note that all the shown strategies show behavior consistent with our characterization of partners: 
If a co-player fully cooperated in the previous $n$ rounds, these strategies prescribe to continue with cooperation. 
If the co-player defected, however, they cooperate with a markedly reduced cooperation probability that satisfies the constraints in Eqs.~\eqref{eq:two_bit_conditions} -- \eqref{eq:counting}. 

In a next step, we systematically explore the impact of three key parameters: the cost-to-benefit ratio~$c/b$, the selection strength $\beta$, and the memory length~$n$. 
In each case, we record how these parameters affect the abundance of partner strategies and the population's average cooperation rate. 
Overall, the effect of each parameter is largely as expected~(\figref{fig:evolutionary_results}\textbf{C,D}).  
In particular, interactions are most cooperative when the cost-to-benefit ratio is small, such that cooperation is cheap. 
This effect is magnified for stronger selection strengths. 
Two results, however, are particularly noteworthy. 
First, the curves representing evolving cooperation rates align with the prevalence of partner strategies. 
This observation suggests that partner strategies are indeed crucial for the evolution of cooperation. 
Second, higher memory only has a notably positive effect on cooperation for reactive-$n$ strategies. 
In contrast, for counting strategies the effect of increasing $n$ is negligible. 
This observation highlights that the timing of cooperation is important, even in additive games such as the donation game.


\section*{Discussion}

Direct reciprocity is a key mechanism for cooperation, based on the intuition that individuals are more likely to cooperate when they meet repeatedly~\citep{nowak:Science:2006}.
To capture the logic of reciprocity, most previous theoretical studies focus on a subset of strategies, the  memory-1 strategies~\citep{nowak:Nature:1993,imhof:PNAS:2005,grujic:jtb:2012,van-segbroeck:prl:2012,press:PNAS:2012,stewart:pnas:2013,Toupo:IJBC:2014,stewart:pnas:2014, akin:EGADS:2016,glynatsi:scientific:2020,chen:PNASnexus:2023}. 
This set is comparably easy to work with: 
the number of deterministic memory-1 strategies is manageable; most strategies are easy to interpret; and payoffs can be computed efficiently~\citep{sigmund2010}. 
At the same time, however, this strategy space leaves out many interesting reciprocal behaviors that are of theoretical or empirical relevance.
For example, already simple behaviors such as Tit-for-Two-Tat~\citep{axelrod:AAAS:1981} are not representable with one-round memory.
This shortcoming is particularly consequential for noisy games, where higher-memory strategies are important~\cite{fudenberg:AER:2012}. 
In such games, individuals often take into account information from previous rounds to make sense of a co-player's defection in the last round. 
That is, the earlier history of play provides an important context to interpret the co-player's last-round behavior. 

To make progress, we consider an easily interpretable set of strategies with higher memory. 
These reactive-$n$ strategies take into account a co-player's moves during the past $n$ rounds. 
They capture the basic idea of conditional cooperation: people are responsive to the previous actions of their interaction partners. 
For reactive-$n$ strategies, we derive a convenient method to characterize all `partner strategies' -- strategies that sustain full cooperation in a Nash equilibrium~\citep{akin:EGADS:2016,Hilbe:GEB:2015}. 
We show that for a reactive-$n$ strategy to be a Nash equilibrium, it is not necessary to check all possible deviations. 
It suffices to only check deviations towards (deterministic) self-reactive-$n$ strategies. 
Self-reactive players are particularly simple to describe. 
They only take into account their own previous moves. 
In particular, the future behavior of a self-reactive player is independent of the co-player. 
We use this insight to characterize the reactive-$n$ partner strategies in the repeated prisoner's diemma.
But the same insight can be applied to other contexts. 
For example, it can be equally used to characterize other Nash equilibria (not only the cooperative ones). 
Similarly, it can be used to characterize the Nash equilibria of other repeated games, such as the snowdrift game~\citep{doebeli:EL:2005} or the volunteer's dilemma~\citep{diekmann:jcr:1985}.  
In this way, some of our technical results represent useful tools to make further progress on the theory of repeated games, similar to Press and Dyson's insight that any memory-1 strategy has a memory-1 best response~\citep{press:PNAS:2012}. 

Especially for small memory lengths, the conditions for partner strategies are intuitive. 
For example, for the donation game with $n\!=\!2$ rounds of memory, we end up with three conditions, see Eq.~\eqref{eq:two_bit_conditions}.
({\it i}) If the co-player cooperated twice, continue to cooperate; 
({\it ii}) If the co-player cooperated once, cooperate with a slightly reduced probability of $1\!-\!c/(2b)$ on average. 
({\it iii}) If the co-player did not cooperate at all, reduce the cooperation probability even further, to $1\!-\!c/b$. 
As we increase the memory length to $n\!\ge\!3$, or as we consider more general games, there are more conditions to satisfy, and the conditions become harder to interpret. 
However, the three simple conditions above do generalize to larger $n$ if we focus on the subset of counting strategies. 
These are the reactive-$n$ strategies that merely count how often the co-player cooperated during the last $n$ rounds. 
For counting strategies, we show that for each defection of the co-player in memory, a partner reduces its cooperation probability by $c/(nb)$. 
A partner's generosity decreases in proportion to their opponent's selfishness. 

With respect to {\it sustaining} cooperation, counting strategies thus seem to be just as effective as the more complex reactive-$n$ strategies.
With respect to the {\it evolution} of cooperation, however, they seem seem far less effective. 
In simulations, memory size only has a positive impact on evolving cooperation rates for reactive-$n$ strategies, but not for counting strategies~(\figref{fig:evolutionary_results}). 
These results suggest that memory is not only important to record {\it how often} a co-player cooperated, but also {\it when}. 
Overall, these results shed an important light on the logic of reciprocity for individuals with plausible cognitive abilities. 
While in practice, people's cooperative decisions often depend on the outcome of their last encounter, they rarely depend on that last encounter {\it only}. 
Our results suggest a way how individuals can integrate information from previous interactions to cooperate most effectively.


\subsection*{Materials and Methods}\label{section:materials_and_methods}

Our study combines two independent approaches, an equilibrium analysis and evolutionary simulations.\\

\noindent
{\bf Equilibrium analysis.} Here we only summarize our equilibrium analysis; all details are in the \SI. 
There, we formally introduce the three relevant strategy spaces, memory-$n$ strategies, reactive-$n$ strategies, and self-reactive-$n$ strategies. 
Then we provide an explicit algorithm for computing these strategies' payoffs. 
This algorithm uses a Markov chain approach. 
The states of the Markov chain are the possible combinations of $n$-histories of the two players. 
Given the players' current $n$-histories and their strategies, we can compute the likelihood of observing each possible state one round later. 

In a second step, we explore the partner strategies among the reactive-$n$ strategies. 
To this end, we first generalize some well-known reactive-1 partner strategies: Tit-for-Tat~\citep{axelrod:AAAS:1981} and Generous Tit-for-Tat~\citep{nowak:Nature:1992,molander:jcr:1985}. 
In a next step, we derive a general algorithm to check whether a given reactive-$n$ strategy is a partner. 
We use this algorithm to characterize all reactive-$n$ partners for $n\!\in\!\{1,2,3\}$, for both the donation game and the prisoner's dilemma. 
For counting strategies, we characterize the partners for all $n$.\\

\noindent
{\bf Evolutionary analysis.} For our simulations, we consider a population of size \(N\) where initially all
members are of the same strategy. 
In our case the initial population consists of unconditional defectors. 
In each elementary time step, one individual switches to a new mutant strategy. 
The mutant strategy is generated by randomly drawing cooperation probabilities from the unit interval \([0,1]^{2^n}\). 
If the mutant strategy yields a payoff of \(\pi_{M, k}\), where \(k\) is the number of
mutants in the population, and if residents get a payoff of \(\pi_{R,
k}\), then the fixation probability \(\phi_{M}\) of the mutant strategy can be
calculated explicitly~\citep{nowak:Nature:2004},
\begin{equation}\label{eq:fixation_probability}
  \phi_{M} =\Big(1 + \displaystyle \sum_{i=1}^{N - 1} \prod_{j=1}^{i} e^{- \beta (\pi_{M, j} - \pi_{R, i})} \Big)^{-1}.
\end{equation}
The parameter \(\beta \geq 0\) reflects the strength of selection. 
It measures the importance of relative payoff advantages for the evolutionary success of a strategy. 
When \(\beta\) is small, \(\beta \approx 0\), payoffs become irrelevant, and a strategy's fixation probability approaches
\(\phi_{M} \approx 1 / N\). 
The larger the value of \(\beta\), the more strongly the evolutionary process favors the fixation of strategies with a high
payoff.
Depending on \(\phi_{M}\), the mutant either fixes (becomes the new resident) or goes extinct. 
Afterwards, another mutant strategy is introduced to the population. 
We iterate this elementary population updating process for a large number of
mutant strategies. At each step, we record the current resident
strategy and the resulting average cooperation rate, indicating how often the
resident strategy cooperates with itself. Additionally, we assess how many
resident strategies qualify as partner strategies in our simulation. For a
resident strategy to be classified as a partner, it must satisfy all
inequalities in the respective definition of partner strategies and cooperate
with a probability of at least 95\% after full cooperation.

\section*{Data, Materials, and Software Availability}

The source code used to reproduce the results of this study is available on the
online GitHub repository:
\href{https://github.com/Nikoleta-v3/conditional-cooperation-with-longer-memory}{Nikoleta-v3/conditional-cooperation-with-longer-memory}.
The simulation data have been archived on Zenodo and can be found at:
\href{https://zenodo.org/records/10605988}{zenodo.org/records/10605988}.

{\setlength{\bibsep}{0\baselineskip}
\bibliographystyle{naturemag}
\bibliography{bibliography.bib}
}

\clearpage
\newpage

\clearpage
\newpage



\section*{Supplementary material}

This document provides further details on our methodology and our analytical results. Section 1 summarizes the model. In particular, we introduce all relevant strategy spaces, and we show how to compute long-term payoffs for strategies with more than one-round memory. Section 2 contains our key results. Here, we define partner strategies, we present an algorithm that allows us to verify whether a given reactive-$n$ strategy is a partner, and we apply this algorithm to fully characterize the reactive-$n$ partner strategies for $n=2$ and $n=3$. All proofs are presented in the Appendix in Section 3.

## 1 Model and basic results

### 1.1 The repeated prisoner's dilemma

We consider the infinitely repeated prisoner's dilemma between two players, player 1 and player 2. Each round, each player can either cooperate ($C$) or defect ($D$). The resulting payoffs are given by the matrix

$$\begin{array}{c} \\ C \\ D \end{array} \begin{pmatrix} C & D \\ R & S \\ T & P \end{pmatrix}. \tag{1}$$

Here, $R$ is the reward payoff of mutual cooperation, $T$ is the temptation to defect, $S$ is the sucker's payoff, and $P$ is the punishment payoff for mutual defection. For the game to be a prisoner's dilemma, we require

$$T > R > P > S \quad \text{and} \quad 2R > T+S. \tag{2}$$

That is, mutual cooperation is the best outcome to maximize the players' total payoffs, but each player's dominant action is to defect. For some of our results, we focus on a special case of the prisoner's dilemma, the donation game. This game only depends on two free parameters, the benefit $b$ and the cost $c$ of cooperation. The payoff matrix of the donation game takes the form

$$\begin{array}{c} \\ C \\ D \end{array} \begin{pmatrix} C & D \\ b-c & -c \\ b & 0 \end{pmatrix}. \tag{3}$$

For this game to satisfy the conditions (2) of a prisoner's dilemma, we assume $b>c>0$ throughout.

Players interact in the repeated prisoner's dilemma for infinitely many rounds, and future payoffs are not discounted. A strategy $\sigma^i$ for player $i$ is a rule that tells the player what to do in any given round, depending on the outcome of all previous rounds. Given the player's strategies $\sigma^1$ and $\sigma^2$, one can compute each player $i$'s expected payoff $\pi^i_{\sigma^1,\sigma^2}(t)$ in round $t$. For the entire repeated game, we define the players' payoffs as the expected payoff per round,

$$\pi^i(\sigma^1, \sigma^2) = \lim_{\tau \to \infty} \frac{1}{\tau} \sum_{t=1}^{\tau} \pi^i_{\sigma^1,\sigma^2}(t). \tag{4}$$

For general strategies $\sigma^1$ and $\sigma^2$, the above limit may not always exist. Problems may arise, for example, if one of the players cooperates in the first round, defects in the two subsequent rounds, cooperates in the four rounds thereafter, etc., which prevents the time averages from converging. However, in the following, we focus on strategies with finite memory. When both players adopt such a strategy, the existence of the limit (4) is guaranteed, as we discuss further below.

### 1.2 Finite-memory strategies

In this study, we focus on strategies that ignore all events that happened more than $n$ rounds ago. To define these strategies, we need some notation. An *n-history for player i* is a string $\mathbf{h^i} = (a^i_{-n}, \ldots, a^i_{-1}) \in \{C, D\}^n$. We interpret the string's entry $a^i_{-k}$ as player $i$'s action $k$ rounds ago. We denote the space of all $n$-histories for player $i$ as $H^i$. This space contains $|H^i| = 2^n$ elements. A pair $\mathbf{h} = (\mathbf{h^1}, \mathbf{h^2})$ is called an *n-history of the game*. We use $H = H^1 \times H^2$ to denote the space of all such histories, which contains $|H| = 2^{2n}$ elements.

**Memory-$n$ strategies.** Based on this notation, a *memory-n strategy* for player $i$ as a tuple $\mathbf{m} = (m_\mathbf{h})_{\mathbf{h} \in H}$. Each input $\mathbf{h} = (h^i, h^{-i})$ refers to a possible $n$-history, where now $\mathbf{h^i}$ and $\mathbf{h^{-i}}$ refer to the $n$-histories of the focal player and the co-player, respectively. The corresponding output $m_\mathbf{h} \in [0, 1]$ is the focal player's cooperation probability in the next round, contingent on the outcome of the previous $n$ rounds. We refer to the set of all memory-$n$ strategies as

$$\mathcal{M}_n := \left\{ \mathbf{m} = (m_\mathbf{h})_{\mathbf{h} \in H} \,\middle|\, 0 \leq m_\mathbf{h} \leq 1 \text{ for all } \mathbf{h} \in H \right\} = [0, 1]^{2^{2n}}. \tag{5}$$

This definition leaves the strategy's actions during the first $n$ rounds unspecified, for which no complete $n$-history is yet available. However, because we consider infinitely repeated games without discounting, these first $n$ rounds are usually irrelevant for the long-run dynamics, as we show further below. In the following, we therefore only specify a strategy's move during the first $n$ rounds when necessary.

Among all memory-$n$ spaces $\mathcal{M}_n$, the one with $n = 1$ is the most frequently studied. Memory-1 strategies take the form $\mathbf{m} = (m_{CC}, m_{CD}, m_{DC}, m_{DD})$. The first index refers to the focal player's last action (1-history) and the second index refers to the co-player's last action. As an example of a well-known memory-1 strategy, we mention Win-Stay Lose-Shift [1], $\mathbf{m} = (1, 0, 0, 1)$. However, there are many others [2].

**Reactive-$n$ strategies.** For our following analysis, two particular subsets of memory-$n$ strategies will play an important role. The first subset is the set of *reactive-n strategies*,

$$\mathcal{R}_n := \left\{ \mathbf{m} \in \mathcal{M}_n \,\middle|\, m_{(\mathbf{h^i}, \mathbf{h^{-i}})} = m_{(\tilde{\mathbf{h}}^\mathbf{i}, \mathbf{h^{-i}})} \text{ for all } \mathbf{h^i}, \tilde{\mathbf{h}}^\mathbf{i} \in H^i \text{ and } \mathbf{h^{-i}} \in H^{-i} \right\}. \tag{6}$$

That is, reactive-$n$ strategies are independent of the focal player's own $n$-history. The space of reactive-$n$ strategies can be naturally identified with the space of all $2^n$-dimensional vectors

$$\mathbf{p} = (p_{\mathbf{h^{-i}}})_{\mathbf{h^{-i}} \in H^{-i}} \text{ with } 0 \leq p_{\mathbf{h^{-i}}} \leq 1 \text{ for all } \mathbf{h^{-i}} \in H^{-i}. \tag{7}$$

In this reduced representation, each entry $p_{\mathbf{h}^{-\mathbf{i}}}$ corresponds to the player's cooperation probability in the next round based on the co-player's actions in the previous $n$ rounds. Again, the most studied case of reactive-$n$ strategies is when $n=1$. Here, the reduced representation according to Eq. (7) takes the form $\mathbf{p}=(p_C, p_D)$. Probably the best-known example of a reactive-1 strategy is Tit-for-Tat, TFT [3]. TFT cooperates if and only if the co-player cooperated in the previous round. Hence, its memory-1 representation is $\mathbf{m}=(1,0,1,0)$, whereas its reduced representation is $\mathbf{p}=(1,0)$. Another example is the strategy Generous Tit-for-Tat, GTFT [4, 5]. GTFT occasionally cooperates even if the co-player defected. For that strategy, the memory-1 representation is $\mathbf{m}=(1, p_D^*, 1, p_D^*)$, and the reduced representation is $\mathbf{p}=(1, p_D^*)$, where

$$p_D^* := \min\{1-(T-R)/(R-S),\ (R-P)/(T-P)\}. \tag{8}$$

In the special case that payoffs are given by the donation game, this condition simplifies to $p_D^* = 1 - c/b$.

**Self-reactive-$n$ strategies.** The other important subspace of memory-$n$ strategies is the set of self-reactive-$n$ strategies,

$$\mathcal{S}_n := \left\{ \mathbf{m} \in \mathcal{M}_n \ \Big|\ m_{(\mathbf{h}^\mathbf{i}, \mathbf{h}^{-\mathbf{i}})} = m_{(\mathbf{h}^\mathbf{i}, \tilde{\mathbf{h}}^{-\mathbf{i}})} \text{ for all } \mathbf{h}^\mathbf{i} \in H^i \text{ and } \mathbf{h}^{-\mathbf{i}}, \tilde{\mathbf{h}}^{-\mathbf{i}} \in H^{-i} \right\}. \tag{9}$$

These strategies only depend on the focal player's own decisions during the last $n$ rounds, independent of the co-player's decisions. Again, we can identify any self-reactive-$n$ strategies with a $2^n$-dimensional vector,

$$\tilde{\mathbf{p}} = (\tilde{p}_{\mathbf{h}^\mathbf{i}})_{\mathbf{h}^\mathbf{i} \in H^i} \text{ with } 0 \leq \tilde{p}_{\mathbf{h}^\mathbf{i}} \leq 1 \text{ for all } \mathbf{h}^\mathbf{i} \in H^i. \tag{10}$$

Each entry $\tilde{p}_{\mathbf{h}^\mathbf{i}}$ corresponds to the player's cooperation probability in the next round, contingent on the player's own actions in the previous $n$ rounds. A special subset of self-reactive strategies is given by the round-$k$-repeat strategies, for some $1 \leq k \leq n$. In any given round, players with a *round-$k$-repeat strategy* $\tilde{\mathbf{p}}^{k-\text{Rep}}$ choose the same action as they did $k$ rounds ago. Formally, the entries of $\tilde{\mathbf{p}}^{k-\text{Rep}}$ are defined by

$$p_{\mathbf{h}^\mathbf{i}}^{k-\text{Rep}} = \begin{cases} 1 & \text{if } a_{-k}^i = C \\ 0 & \text{if } a_{-k}^i = D. \end{cases} \tag{11}$$

From this point forward, we will use the notations $\mathbf{m}$, $\mathbf{p}$, and $\tilde{\mathbf{p}}$ to denote memory-$n$, reactive-$n$, and self-reactive-$n$ strategies, respectively. We say these strategies are *pure* or *deterministic* if all conditional cooperation probabilities are either zero or one. If all cooperation probabilities are strictly between zero and one, we say the strategy is *strictly stochastic*. When it is convenient to represent the self-reactive repeat strategies as elements of the memory-$n$ strategy space, we write $\mathbf{m}^{k-\text{Rep}} \in [0,1]^{2^{2n}}$ instead of $\tilde{\mathbf{p}}^{k-\text{Rep}} \in [0,1]^{2^n}$.

### 1.3 Computing the payoffs of finite-memory strategies

**A Markov chain representation.** The interaction between two players with memory-$n$ strategies $\mathbf{m}^1$ and $\mathbf{m}^2$ can be represented as a Markov chain. The states of the Markov chain are the possible $n$-histories $\mathbf{h} \in H$. To compute the transition probabilities from one state to another within a single round, suppose players

4

currently have the $n$-history $\mathbf{h}=(\mathbf{h^1},\mathbf{h^2})$ in memory. Then the transition probability that the state after one round is $\tilde{\mathbf{h}}=(\tilde{\mathbf{h}}^1,\tilde{\mathbf{h}}^2)$ is a product of two factors,

$$M_{\mathbf{h},\tilde{\mathbf{h}}} = x^1 \cdot x^2, \tag{12}$$

The two factors represent the (independent) decisions of the two players,

$$x^i = \begin{cases} m^i_{(\mathbf{h^i},\mathbf{h^{-i}})} & \text{if } \tilde{a}^i_{-1}=C, \text{ and } \tilde{a}^i_{-t}=a^i_{-t+1} \text{ for } t\in\{2,\ldots,n\} \\ 1-m^i_{(\mathbf{h^i},\mathbf{h^{-i}})} & \text{if } \tilde{a}^i_{-1}=D, \text{ and } \tilde{a}^i_{-t}=a^i_{-t+1} \text{ for } t\in\{2,\ldots,n\} \\ 0 & \text{if } \tilde{a}^i_{-t} \neq a^i_{-t+1} \text{ for some } t\in\{2,\ldots,n\}. \end{cases} \tag{13}$$

The resulting $2^{2n} \times 2^{2n}$ transition matrix $M=(M_{\mathbf{h},\tilde{\mathbf{h}}})$ fully describes the dynamics among the two players after the first $n$ rounds. More specifically, suppose $\mathbf{v}(t) = (v_\mathbf{h}(t))_{\mathbf{h}\in H}$ is the probability distribution of observing state $\mathbf{h}$ after players made their decisions for round $t \geq n$. Then the respective probability distribution after round $t+1$ is given by $\mathbf{v}(t+1) = \mathbf{v}(t) \cdot M$. The long-run dynamics is particularly simple to describe when the matrix $M$ is primitive (which happens, for example, when the two strategies $m^i_\mathbf{h}$ are strictly stochastic). In that case, it follows by the theorem of Perron and Frobenius that $\mathbf{v}(t)$ converges to some $\mathbf{v}$ as $t \to \infty$. As a result, also the respective time average exists and converges to $\mathbf{v}$,

$$\mathbf{v} = \lim_{\tau\to\infty} \frac{1}{\tau} \sum_{t=n}^{n+\tau-1} \mathbf{v}(t). \tag{14}$$

This limiting distribution $\mathbf{v}$ can be computed as the unique solution of the system $\mathbf{v}=\mathbf{v}M$, with the additional constraint that the entries of $\mathbf{v}$ need to sum up to one.

But even when $M$ is not ergodic, $\mathbf{v}(t)$ still converges to an invariant distribution $\mathbf{v}$ that satisfies $\mathbf{v}=\mathbf{v}M$. However, in that case, the system $\mathbf{v}=\mathbf{v}M$ no longer has a unique solution. Instead, the limiting distribution $\mathbf{v}$ depends on the very first $n$-history after the first $n$ rounds, $\mathbf{v}(n)$, which in turn depends on the players' moves during the first $n$ rounds.

**A formula for the payoffs among memory-$n$ players.** Based on the above considerations, we can derive an explicit formula for the payoffs according to Eq. (4) when players use memory-$n$ strategies $\mathbf{m^1}$ and $\mathbf{m^2}$. To this end, we introduce a $2^{2n}$-dimensional vector $\mathbf{g^i}(k) = (g^i_\mathbf{h}(k))_{\mathbf{h}\in H}$, that takes an $n$-history $\mathbf{h}$ as an input and returns player $i$'s payoff $k$ rounds ago, for $k \leq n$. That is,

$$g^i_\mathbf{h}(k) = \begin{cases} R & \text{if } a^i_{-k}=C \text{ and } a^{-i}_{-k}=C \\ S & \text{if } a^i_{-k}=C \text{ and } a^{-i}_{-k}=D \\ T & \text{if } a^i_{-k}=D \text{ and } a^{-i}_{-k}=C \\ P & \text{if } a^i_{-k}=D \text{ and } a^{-i}_{-k}=D. \end{cases} \tag{15}$$

Now for a given $t \geq n$, given that $\mathbf{v}(t)$ captures the state of the system after round $t$, we can write player $i$'s

expected payoff in that round as

$$\pi^i_{\mathbf{m^1},\mathbf{m^2}}(t) = \langle \mathbf{v}(t), \mathbf{g^i}(1) \rangle = \sum_{\mathbf{h}\in H} v_{\mathbf{h}}(t) \cdot g^i_{\mathbf{h}}(1). \tag{16}$$

As a result, we obtain for the player's average payoff across all rounds

$$\begin{aligned}
\pi^i(\mathbf{m^1},\mathbf{m^2}) &\stackrel{(4)}{=} \lim_{\tau\to\infty} \frac{1}{\tau}\sum_{t=1}^{\tau} \pi^i_{\mathbf{m^1},\mathbf{m^2}}(t) = \lim_{\tau\to\infty} \frac{1}{\tau}\sum_{t=n}^{n+\tau-1} \pi^i_{\mathbf{m^1},\mathbf{m^2}}(t) \\
&\stackrel{(16)}{=} \lim_{\tau\to\infty} \frac{1}{\tau}\sum_{t=n}^{n+\tau-1} \langle \mathbf{v}(t), \mathbf{g^i}(1) \rangle = \Big\langle \lim_{\tau\to\infty} \frac{1}{\tau}\sum_{t=n}^{n+\tau-1} \mathbf{v}(t), \mathbf{g^i}(1) \Big\rangle \\
&\stackrel{(14)}{=} \langle \mathbf{v}, \mathbf{g^i}(1) \rangle.
\end{aligned} \tag{17}$$

That is, given we know the invariant distribution $\mathbf{v}$ that captures the game's long-run dynamics, it is straightforward to compute payoffs by taking the scalar product with the vector $\mathbf{g^i}(1)$. With a similar approach as in Eq. (17), one can also show

$$\langle \mathbf{v}, \mathbf{g^i}(1) \rangle = \langle \mathbf{v}, \mathbf{g^i}(2) \rangle = \ldots = \langle \mathbf{v}, \mathbf{g^i}(n) \rangle. \tag{18}$$

That is, to compute player $i$'s expected payoff, it does not matter whether one refers to the last round of an $n$-history or to an earlier round of an $n$-history. All rounds $k$ with $1 \le k \le n$ are equivalent.

### 1.4 An Extension of Akin's Lemma

The above Markov chain approach allows us to analyze games when both players adopt memory-$n$ strategies. But even if only one player adopts a memory-$n$ strategy (and the other player's strategy is arbitrary), one can still derive certain constraints on the game's long-run dynamics. One such constraint was first described by Akin [6]: if player 1 adopts a memory-1 strategy $\mathbf{m}$ against an arbitrary opponent, and if the time average $\mathbf{v}$ defined by the right hand side of Eq. (14) exists, then

$$\langle \mathbf{v}, \mathbf{m} - \mathbf{m}^{1-\text{Rep}} \rangle = 0. \tag{19}$$

That is, the limiting distribution $\mathbf{v}$ needs to be orthogonal to the vector $\mathbf{m} - \mathbf{m}^{1-\text{Rep}}$. This result has been termed *Akin's Lemma* [7]. With similar methods as in Ref. [6], one can generalize this result to the context of memory-$n$ strategies.

**Lemma 1** (A generalized version of Akin's Lemma)
*Let player 1 use a memory-$n$ strategy, and let player 2 use an arbitrary strategy. For the resulting game and all $t \ge n$, let $\mathbf{v}(t) = (v_{\mathbf{h}}(t))_{\mathbf{h}\in H}$ denote the probability distribution of observing each possible $n$-history $\mathbf{h} \in H$ after players made their decisions for round $t$. Moreover, suppose the respective time average $\mathbf{v}$*

*according to Eq. (14) exists. Then for each k with $1 \leq k \leq n$, we obtain*

$$\langle \mathbf{v}, \mathbf{m} - \mathbf{m}^{k-Rep} \rangle = 0. \tag{20}$$

All proofs are presented in the Appendix. Here we provide an intuition. The expression $\langle \mathbf{v}, \mathbf{m} \rangle = \sum_{\mathbf{h}} v_{\mathbf{h}} m_{\mathbf{h}}$ can be interpreted as player 1's average cooperation rate across all rounds of the repeated game. To compute that average cooperation rate, one first draws an $n$-history $\mathbf{h}$ (with probability $v_{\mathbf{h}}$), and then one computes how likely player 1 would cooperate in the subsequent round (with probability $m_{\mathbf{h}}$). Alternatively, one could compute the average cooperation rate by drawing an $n$-history $\mathbf{h}$ and then checking how likely player 1 was to cooperate $k$ rounds ago, according to that $n$-history. That second interpretation leads to the expression $\langle \mathbf{v}, \mathbf{m}^{k-\mathrm{Rep}} \rangle$. According to Eq. (20), both interpretations are equivalent.

## 2 Characterizing the partner strategies among the reactive-$n$ strategies

### 2.1 Partner strategies

In this study, we are interested in identifying strategies that can sustain full cooperation in a Nash equilibrium. Strategies with these properties have been termed as being of *Nash type* by Akin [6], or as *partner strategies* by Hilbe *et al* [8]. In the following, we formally define them.

**Definition** (Partner strategies)

(i) A strategy $\sigma$ for the repeated prisoner's dilemma is a *Nash equilibrium* if it is a best response to itself. That is, we require $\pi^1(\sigma, \sigma)$ to exist and

$$\pi^1(\sigma, \sigma) \geq \pi^1(\sigma', \sigma) \text{ for all other strategies } \sigma' \text{ for which } \pi^1(\sigma', \sigma) \text{ exists.} \tag{21}$$

(ii) A player's strategy is *nice*, if the player is never the first to defect.
(iii) A *partner strategy* is a strategy that is both nice and a Nash equilibrium.

Several remarks are in order. First, we note that when two players with nice strategies interact, they both cooperate in every round. Partner strategies thus sustain mutual cooperation in a Nash equilibrium. Second, if a memory-$n$ strategy $\mathbf{m} = (m_{\mathbf{h}})_{\mathbf{h} \in H}$ is to be nice, it needs to cooperate after $n$ rounds of mutual cooperation. In other words, if $\mathbf{h_C} = (\mathbf{h_C^i}, \mathbf{h_C^{-i}})$ is the $n$-history that consists of mutual cooperation for the past $n$ rounds, then the strategy needs to respond by cooperating with certainty, $m_{\mathbf{h_C}} = 1$. Similarly, a nice reactive-$n$ strategy needs to satisfy $p_{\mathbf{h_C^{-i}}} = 1$. Third, we note that our definition of Nash equilibria only requires that players cannot profitably deviate towards strategies *for which a payoff can be defined*. If the strategy $\sigma$ is a memory-$n$ strategy, in the following we make the slightly looser requirement that the strategy is a best response among all $\sigma'$ for which the limit (14) exists. Fourth, in general it is a difficult task to verify that any given strategy $\sigma$ is a Nash equilibrium. After all, one needs to verify that it yields the highest payoff according to Eq. (21) among all (uncountably) many alternative strategies $\sigma'$. Fortunately, the situation is

somewhat simpler if the strategy under consideration is a memory-$n$ strategy. In that case, it follows from an argument by Press and Dyson [9] that one only needs to compare the strategy to all other memory-$n$ strategies. However, this still leaves us with uncountably many strategies to check. In fact, it is one aim of this paper to show that for reactive-$n$ strategies, it suffices to check finitely many alternative strategies.

## 2.2 Tit For Tat and Generous Tit For Tat with arbitrary memory lengths

**Zero-determinant strategies with $n$ rounds memory.** Before we provide a general algorithm to identify reactive-$n$ partner strategies, we first generalize some of the well-known reactive-1 partner strategies, `TFT` and `GTFT`, to the case of memory-$n$. To this end, we use Lemma 1 to develop a theory of zero-determinant strategies within the class of memory-$n$ strategies, see also Refs. [10, 11]. In the following, we say a memory-$n$ strategy $\mathbf{m}$ is a *zero-determinant strategy* if there are integers $k_1, k_2, k_3 \leq n$ and real numbers $\alpha, \beta, \gamma$ such that $\mathbf{m^i}$ can be written as

$$\mathbf{m^i} = \alpha\, \mathbf{g}^i(k_1) + \beta\, \mathbf{g}^{-i}(k_2) + \gamma\, \mathbf{1} + \mathbf{m}^{k_3-\text{Rep}}. \tag{22}$$

In this expression, $\mathbf{g}^i(k)$ is the vector that returns player $i$'s payoff $k$ rounds ago, as defined by Eq. (15), $\mathbf{m}^{k-\text{Rep}}$ is the memory-$n$ strategy that repeats player $i$'s own move $k$ rounds ago, and $\mathbf{1}$ is the $2^{2n}$-dimensional vector for which every entry is one. Using the generalized version of Akin's Lemma, we obtain

$$\begin{aligned}
0 &\stackrel{(20)}{=} \langle \mathbf{v}, \mathbf{m} - \mathbf{m}^{k_3-\text{Rep}} \rangle \\
&\stackrel{(22)}{=} \langle \mathbf{v}, \alpha \mathbf{g}^i(k_1) + \beta \mathbf{g}^{-i}(k_2) + \gamma \mathbf{1} \rangle \\
&= \alpha \langle \mathbf{v}, \mathbf{g}^i(k_1) \rangle + \beta \langle \mathbf{v}, \mathbf{g}^{-i}(k_2) \rangle + \gamma \langle \mathbf{v}, \mathbf{1} \rangle \\
&\stackrel{(17),(18)}{=} \alpha\, \pi^i(\mathbf{m^i}, \sigma^{-i}) + \beta\, \pi^{-i}(\mathbf{m^i}, \sigma^{-i}) + \gamma.
\end{aligned} \tag{23}$$

That is, a player with a zero-determinant strategy enforces a linear relationship between the players' payoffs, irrespective of the co-player's strategy. Remarkably, the parameters $\alpha$, $\beta$, and $\gamma$ of that linear relationship are entirely under player $i$'s control.

**Generalized versions of Tit-for-tat.** One interesting special case arises if $k_1 = k_2 = k_3 =: k$ and $\alpha = -\beta = 1/(T-S)$, $\gamma = 0$. In that case, formula (22) yields the strategy with entries

$$m_\mathbf{h} = \begin{cases} 1 & \text{if } a^{-i}_{-k} = C \\ 0 & \text{if } a^{-i}_{-k} = D \end{cases}$$

Therefore, a player with that strategy cooperates if and only if the co-player cooperated $k$ rounds ago. Thus, the strategy implements `TFT` (for $k=1$) or delayed versions thereof (for $k>1$). By Eq. (23), the strategy enforces equal payoffs against any co-player,

$$\pi^i(\mathbf{m^i}, \sigma^{-i}) = \pi^{-i}(\mathbf{m^i}, \sigma^{-i}). \tag{24}$$

8

Moreover, this strategy is nice if we additionally require it to unconditionally cooperate during the first $k$ rounds. Given this additional requirement, the payoff of $\mathbf{m^i}$ against itself is $R$. Moreover, the strategy is a Nash equilibrium. To see why, suppose to the contrary that there is a strategy $\sigma^{-i}$ with $\pi^{-i}(\mathbf{m^i}, \sigma^{-i}) > R$. Then it follows from (24) that $\pi^i(\mathbf{m^i}, \sigma^{-i}) + \pi^{-i}(\mathbf{m^i}, \sigma^{-i}) > 2R$. That is, the total payoff per round exceeds $2R$, which is incompatible with the basic assumptions on a prisoner's dilemma, Eq. (2). We conclude that all these versions of TFT are nice and they are Nash equilibria. Hence, they are partner strategies.

**Generalized versions of Generous Tit-for-Tat.** Another interesting special case arises in the donation game if $k_1 = k_2 = k_3 =: k$ and $\alpha = 0$, $\beta = -1/b$, $\gamma = 1 - c/b$. In that case Eq. (22) yields the strategy with entries

$$m_\mathbf{h} = \begin{cases} 1 & \text{if } a^{-i}_{-k} = C \\ 1 - c/b & \text{if } a^{-i}_{-k} = D \end{cases}$$

That is, the generated strategy is GTFT (if $k = 1$), or a delayed version thereof (for $k > 1$). By Eq. (23), the enforced payoff relationship is $\pi^{-i}(\mathbf{m^i}, \sigma^{-i}) = b - c$. That is, the co-player always obtains the mutual cooperation payoff, irrespective of the co-player's strategy. In particular, all these versions of GTFT are Nash equilibria (independent of how they act during the first $n$ rounds). If we additionally require them to cooperate during the first $n$ rounds, they are also nice. Hence, they are partner strategies.

### 2.3 An algorithm to check whether a reactive-$n$ strategy is a Nash equilibrium

**Sufficiency of checking pure self-reactive strategies.** After discussing these particular cases, we would like to derive a general algorithm that allows us to verify whether a given reactive-$n$ strategy is a Nash equilibrium. In principle, this requires us to check the payoff of any other strategy (including strategies that have a much longer memory length than $n$). Fortunately, however, some simplifications are possible when we use an insight by Press and Dyson [9]. They discussed the case where one player uses a memory-1 strategy and the other player employs a longer memory strategy. They demonstrated that the payoff of the player with the longer memory is exactly the same as if the player had employed a specific shorter-memory strategy, disregarding any history beyond what is shared with the short-memory player. Here we show a result that follows a similar intuition. If there is a part of the game's history that one player does not take into account, then the co-player gains nothing by considering that part of the history.

**Lemma 2** (Against reactive strategies, any feasible payoff can be generated with self-reactive strategies)
*Let $\mathbf{p} \in \mathcal{R}_n$ be a reactive strategy for player 2. Moreover, suppose player 1 adopts some strategy $\sigma$ such that for the resulting game, the time average $\mathbf{v}$ according to Eq. (14) exists. Then there is a self-reactive-$n$ strategy $\tilde{\mathbf{p}} \in \mathcal{S}_n$ such that $\pi^i(\sigma, \mathbf{p}) = \pi^i(\tilde{\mathbf{p}}, \mathbf{p})$ for $i \in \{1, 2\}$.*

For an illustration of this result, see Figure S1. It shows that against a reactive-2 player, any payoff that can be achieved with a memory-2 strategy can already be achieved with a self-reactive-2 strategy.

If we are to verify that some given reactive-$n$ strategy $\mathbf{p}$ is a Nash equilibrium, Lemma 2 simplifies our task considerably. Instead of checking condition (21) for all possible strategies $\sigma'$, we only need to check it

9

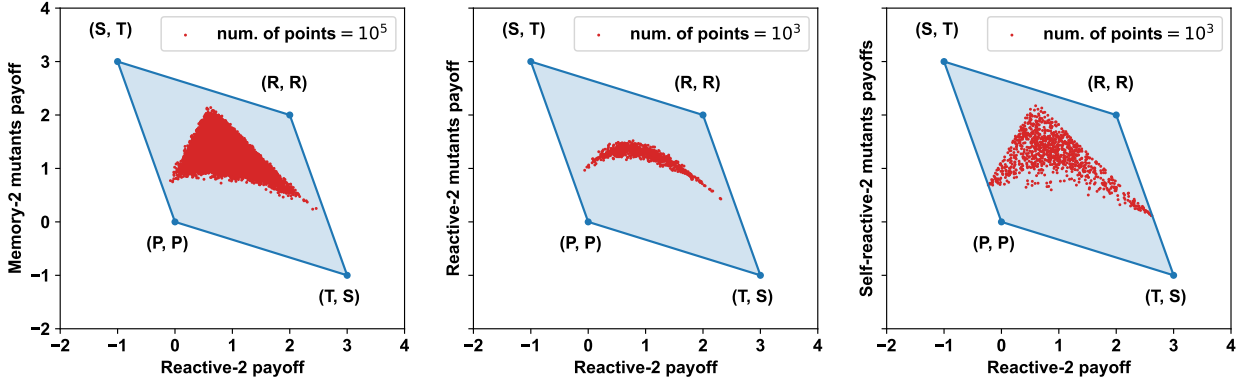

**Figure S1: Feasible payoffs for a reactive-2 strategy.** We consider a player with reactive-2 strategy $\mathbf{p} = (0.37, 0.89, 0.95, 0.23)$. The player interacts with many other players (referred to as 'mutants') who adopt either some random memory-2 strategy (left), a random reactive-2 strategy (middle), or a random self-reactive-2 strategy (right panel). The panels show the resulting payoffs to the two players as red dots, with the $x$-axis showing the payoff of the focal player, and the $y$-axis showing the payoff of the mutants. We observe that when mutants use memory-2 strategies and self-reactive-2 strategies, we obtain the same region of feasible payoffs, in line with Lemma 2. In contrast, if mutants are restricted to reactive-2 strategies, the set of feasible payoffs is strictly smaller. Here, we consider a donation game with $b = 3$ and $c = 1$.

for all self-reactive strategies $\tilde{\mathbf{p}} \in \mathcal{S}_n$. The following result simplifies our task even further.

**Theorem 1** (To any reactive strategy, there is a best response among the pure self-reactive strategies)
*For any reactive strategy $\mathbf{p} \in \mathcal{R}_n$ there is some pure self-reactive strategy $\tilde{\mathbf{p}} \in \mathcal{S}_n$ such that*

$$\pi^1(\tilde{\mathbf{p}}, \mathbf{p}) \geq \pi^1(\sigma', \mathbf{p}) \text{ for all other strategies } \sigma' \text{ for which the limit (14) exists.} \tag{25}$$

This result implies that we only need to check finitely many other strategies if we are to verify that some given reactive-$n$ strategy is a Nash equilibrium.

**Corollary 1** (An algorithm to check whether a reactive-$n$ strategy is a Nash equilibrum)
*A reactive strategy $\mathbf{p} \in \mathcal{R}_n$ is a Nash equilibrium if and only if $\pi^1(\mathbf{p}, \mathbf{p}) \geq \pi^1(\tilde{\mathbf{p}}, \mathbf{p})$ for all pure self-reactive strategies $\tilde{\mathbf{p}} \in \mathcal{S}_n$.*

Corollary 1 gives us a straightforward procedure to check whether a given reactive strategy $\mathbf{p}$ is a Nash equilibrium (for a depiction, see Algorithm 1). To verify that $\mathbf{p}$ is a Nash equilibrium, we merely need to compare its payoff against itself to the payoff of a deviation towards one of the $2^n$ pure self-reactive strategies.

**A more efficient way to calculate payoffs.** For the remainder of this section, we thus assume that player 1 uses a self-reactive-$n$ strategy $\tilde{\mathbf{p}} = (\tilde{p}_{\mathbf{h}^i})_{\mathbf{h}^i \in H^i}$, whereas player 2 uses a reactive-$n$ strategy $\mathbf{p} = (p_{\mathbf{h}^{-i}})_{\mathbf{h}^{-i} \in H^{-i}}$. Our algorithm to compute payoffs for the two players in Section 1.3 would require us to interpret the two strategies as memory-$n$ strategies. We would thus compute a left eigenvector of a $2^{2n} \times 2^{2n}$ transition matrix. In the following, however, we show that for games between reactive and self-reactive players, it suffices to consider a $2^n \times 2^n$ transition matrix. This efficiency gain is possible because both players only consider

10

**Algorithm 1:** An algorithm to verify whether a given reactive strategy **p** is a Nash equilibrium.

   **input:** $\mathbf{p}, n$

   pure_self_reactive_strategies $\leftarrow \{\tilde{\mathbf{p}} \mid \tilde{\mathbf{p}} \in \{0,1\}^{2^n}\}$ ;

   isNash $\leftarrow$ True ;

   **for** $\tilde{\mathbf{p}} \in$ *pure_self_reactive_strategies* **do**

      **if** $\pi^1(\mathbf{p},\mathbf{p}) < \pi^1(\tilde{\mathbf{p}},\mathbf{p})$ **then**

         isNash $\leftarrow$ False ;

   **return** (**p**, isNash) ;

---

player 1's past actions. Instead of taking the space of all of the game's $n$-histories $H = H^1 \times H^2$ as the state space, we can thus take the space $H^1$. Let $h^1 = (a^1_{-n}, \ldots, a^1_{-1})$ be the state in the current round. Then we obtain the following probability that the state after one round is $\tilde{h}^1 = (\tilde{a}^1_{-n}, \ldots, \tilde{a}^1_{-1})$,

$$\tilde{M}_{\mathbf{h^1},\tilde{\mathbf{h}}^1} = \begin{cases} \tilde{p}_{\mathbf{h^1}} & \text{if } \tilde{a}^1_{-1} = C, \text{ and } \tilde{a}^1_{-t} = a^1_{-t+1} \text{ for all } t \in \{2, \ldots, n\} \\ 1 - \tilde{p}_{\mathbf{h^1}} & \text{if } \tilde{a}^1_{-1} = D, \text{ and } \tilde{a}^1_{-t} = a^1_{-t+1} \text{ for all } t \in \{2, \ldots, n\} \\ 0 & \text{if } \tilde{a}^1_{-t} \neq a^1_{-t+1} \text{ for some } t \in \{2, \ldots, n\}. \end{cases} \quad (26)$$

Similar to the vector **v** for matrix $M$, let $\tilde{\mathbf{v}} = (\tilde{v}_{\mathbf{h^1}})_{\mathbf{h^1} \in H^1}$ be the limiting distribution of the dynamics defined by $\tilde{M}$ (which only in exceptional cases depends on player 1's behavior during the first $n$ rounds). Then the players' payoffs are given by

$$\begin{aligned}
\pi^1(\tilde{\mathbf{p}},\mathbf{p}) &= \sum_{\mathbf{h^1} \in H^1} \tilde{v}_{\mathbf{h^1}} \Big(\tilde{\mathbf{p}}_{\mathbf{h^1}} \mathbf{p}_{\mathbf{h^1}} \cdot R + \tilde{\mathbf{p}}_{\mathbf{h^1}}(1-\mathbf{p}_{\mathbf{h^1}}) \cdot S + (1-\tilde{\mathbf{p}}_{\mathbf{h^1}})\mathbf{p}_{\mathbf{h^1}} \cdot T + (1-\tilde{\mathbf{p}}_{\mathbf{h^1}})(1-\mathbf{p}_{\mathbf{h^1}}) \cdot P\Big), \\
\pi^2(\tilde{\mathbf{p}},\mathbf{p}) &= \sum_{\mathbf{h^1} \in H^1} \tilde{v}_{\mathbf{h^1}} \Big(\tilde{\mathbf{p}}_{\mathbf{h^1}} \mathbf{p}_{\mathbf{h^1}} \cdot R + \tilde{\mathbf{p}}_{\mathbf{h^1}}(1-\mathbf{p}_{\mathbf{h^1}}) \cdot T + (1-\tilde{\mathbf{p}}_{\mathbf{h^1}})\mathbf{p}_{\mathbf{h^1}} \cdot S + (1-\tilde{\mathbf{p}}_{\mathbf{h^1}})(1-\mathbf{p}_{\mathbf{h^1}}) \cdot P\Big).
\end{aligned} \quad (27)$$

**Example: Payoffs and best responses with one-round memory.** To illustrate the above results, we consider the case $n = 1$. Assume player 1's self-reactive strategy is $\tilde{\mathbf{p}}^{\mathbf{1}} = (\tilde{p}^1_C, \tilde{p}^1_D)$ and player 2's reactive strategy is $\mathbf{p^2} = (p^2_C, p^2_D)$. If we use the algorithm in Section 1.3, we first formally represent these strategies as memory-1 strategies, $\mathbf{m^1} = (\tilde{p}^1_C, \tilde{p}^1_C, \tilde{p}^1_D, \tilde{p}^1_D)$ and $\mathbf{m^2} = (p^2_C, p^2_D, p^2_C, p^2_D)$. The respective transition matrix according to Eq. (12) is

$$M = \begin{pmatrix} \tilde{p}^1_C p^2_C & \tilde{p}^1_C(1-p^2_C) & (1-\tilde{p}^1_C)p^2_C & (1-\tilde{p}^1_C)(1-p^2_C) \\ \tilde{p}^1_C p^2_D & \tilde{p}^1_C(1-p^2_D) & (1-\tilde{p}^1_C)p^2_D & (1-\tilde{p}^1_C)(1-p^2_D) \\ \tilde{p}^1_D p^2_C & \tilde{p}^1_D(1-p^2_C) & (1-\tilde{p}^1_D)p^2_C & (1-\tilde{p}^1_D)(1-p^2_C) \\ \tilde{p}^1_D p^2_D & \tilde{p}^1_D(1-p^2_D) & (1-\tilde{p}^1_D)p^2_D & (1-\tilde{p}^1_D)(1-p^2_D) \end{pmatrix}. \quad (28)$$

Assuming player 1's strategy is different from the one-round repeat strategy, $\tilde{\mathbf{p}}^{\mathbf{1}} \neq (1,0)$, this transition matrix has a unique invariant distribution,

$$\mathbf{v} = \left( \frac{\tilde{p}_D^1\left(\tilde{p}_C^1(p_C^2-p_D^2)+p_D^2\right)}{1-(\tilde{p}_C^1-\tilde{p}_D^1)},\ \frac{\tilde{p}_D^1\left(1-\tilde{p}_C^1(p_C^2-p_D^2)-p_D^2\right)}{1-(\tilde{p}_C^1-\tilde{p}_D^1)},\ \frac{(1-\tilde{p}_C^1)\left(\tilde{p}_D^1(p_C^2-p_D^2)+p_D^2\right)}{1-(\tilde{p}_C^1-\tilde{p}_D^1)},\ \frac{(1-\tilde{p}_C^1)\left(1-\tilde{p}_D^1(p_C^2-p_D^2)-p_D^2\right)}{1-(\tilde{p}_C^1-\tilde{p}_D^1)} \right).$$

According to Eq. (16), Player 1's payoff is the scalar product

$$\pi^1(\tilde{\mathbf{p}}^1, \mathbf{p}^2) = \langle \mathbf{v}, (R, S, T, P) \rangle. \tag{29}$$

Following Corollary 1, we can use these observations to characterize under which conditions a nice reactive strategy $\mathbf{p}^2 = (1, p_D^2)$ is a partner. To this end, we compute player 1's payoff for all pure self-reactive strategies $\tilde{\mathbf{p}}^1 = (\tilde{p}_C^1, \tilde{p}_D^1)$. These are ALLC $=(1,1)$, ALLD $=(0,0)$, and Alternator $=(0,1)$; we can ignore the one-round repeat strategy (1,0), because depending on the strategy's first round-behavior it is either equivalent to ALLC or to ALLD. The payoffs of these three strategies are

$$\begin{aligned}
\pi^1(\text{ALLC}, \mathbf{p}^2) &= R, \\
\pi^1(\text{ALLD}, \mathbf{p}^2) &= p_D^2 \cdot T + (1-p_D^2) \cdot P \\
\pi^1(\text{Alternator}, \mathbf{p}^2) &= p_D^2/2 \cdot R + (1-p_D^2)/2 \cdot S + 1/2 \cdot T.
\end{aligned} \tag{30}$$

We conclude that player 2's reactive strategy $\mathbf{p}^2$ is a Nash equilibrium (and hence a partner) if none of these three payoffs exceeds the mutual cooperation payoff $R$. This requirement yields the condition

$$p_D^2 \leq \min\left\{1-(T-R)/(R-S),\ (R-P)/(T-P)\right\}. \tag{31}$$

As one may expect, $\mathbf{p}^2$ is a partner if and only if its generosity $p_D^2$ does not exceed the generosity of GTFT, as defined by Eq. (8).

Instead of computing the $4 \times 4$ matrix $M$ in (28), we could also consider the simplified $2 \times 2$ transition matrix (26). Here, the two possible states are $\mathbf{h}^1 \in \{C, D\}$, and hence the matrix is

$$\tilde{M} = \begin{pmatrix} \tilde{p}_C^1 & 1-\tilde{p}_C^1 \\ \tilde{p}_D^1 & 1-\tilde{p}_D^1 \end{pmatrix}. \tag{32}$$

Again, for $\tilde{\mathbf{p}}^1 \neq (1,0)$, this transition matrix has a unique invariant distribution,

$$\tilde{\mathbf{v}} = (\tilde{v}_C, \tilde{v}_D) = \left( \frac{\tilde{p}_D^1}{1-(\tilde{p}_C^1-\tilde{p}_D^1)},\ \frac{1-\tilde{p}_C^1}{1-(\tilde{p}_C^1-\tilde{p}_D^1)} \right). \tag{33}$$

If we take this invariant distribution and compute player 1's payoff according to Eq. (27), we recover the same expression as in Eq. (29), as expected.

### 2.4 Reactive partner strategies in the donation game

Just as in the previous example with $n=1$, we can use the results of the previous section to characterize the partner strategies for reactive-2 and reactive 3-strategies. For simplicity, we first consider the case of the donation game. Results for the general prisoner's dilemma follow in the next section.

**Reactive-2 partner strategies.** We first consider the case $n=2$. The resulting reactive-2 strategies can be represented as a vector $\mathbf{p} = (p_{CC}, p_{CD}, p_{DC}, p_{DD})$. The entries $p_{\mathbf{h}^{-\mathbf{i}}}$ are the player's cooperation probability, depending on the co-player's actions in the previous two rounds, $\mathbf{h}^{-\mathbf{i}} = (a_{-2}^{-i}, a_{-1}^{-i})$. For the strategy to be nice, we require $p_{CC}=1$. Based on Corollary 1, we obtain the following characterization of partners.

**Theorem 2** (Reactive-2 partner strategies in the donation game)
*A nice reactive-2 strategy $\mathbf{p}$, is a partner strategy if and only if its entries satisfy the conditions*

$$p_{CC}=1, \qquad \frac{p_{CD}+p_{DC}}{2} \leq 1 - \frac{1}{2} \cdot \frac{c}{b}, \qquad p_{DD} \leq 1 - \frac{c}{b}. \tag{34}$$

The resulting conditions can be interpreted as follows: For each time a co-player has defected during the past two rounds, the reactive player's cooperation probability needs to decrease by $c/(2b)$. This reduced cooperation probability is sufficient to incentivize the co-player to cooperate. Interestingly, for the strategy to be a partner, the middle condition in (34) suggests that the exact timing of a co-player's defection is irrelevant. As long as *on average*, the respective cooperation probabilities $p_{CD}$ and $p_{DC}$ are below the required threshold $1-c/(2b)$, the strategy is a Nash equilibrium.

The conditions for a partner become even simpler for *reactive-n counting strategies*. To define these strategies, let $|\mathbf{h}^{-\mathbf{i}}|$ denote the number of $C$'s in a given $n$-history of the co-player. We say a reactive-$n$ strategy $\mathbf{p} = (p_{h^{-i}})_{\mathbf{h}^{-\mathbf{i}} \in \mathbf{H}^{-\mathbf{i}}}$ is a counting strategy if

$$|\mathbf{h}^{-i}| = |\tilde{\mathbf{h}}^{-\mathbf{i}}| \quad \Rightarrow \quad p_{\mathbf{h}^{-\mathbf{i}}} = p_{\tilde{\mathbf{h}}^{-\mathbf{i}}}. \tag{35}$$

That is, the reactive player's cooperation probability only depends on the number of cooperative acts during the past $n$ rounds and not on their timing. Such reactive-$n$ counting strategies can be written as $n+1$-dimensional vectors $\mathbf{r} = (r_k)_{k \in \{n,\ldots,1\}}$, where $r_i$ is the player's cooperation probability if the co-player cooperated $k$ times during the past $n$ rounds. In particular, for reactive-2 counting strategies, we associate $r_2 = p_{CC}$, $r_1 = p_{CD} = p_{DC}$, and $r_0 = p_{DD}$. The following characterization of partners among the reactive-2 counting strategies then follows immediately from Theorem 2.

**Corollary 2** (Partners among the reactive-2 counting strategies)
*A nice reactive-2 counting strategy $\mathbf{r} = (r_2, r_1, r_0)$ is a partner strategy if and only if*

$$r_2=1, \qquad r_1 \leq 1 - \frac{1}{2} \cdot \frac{c}{b}, \qquad r_0 \leq 1 - \frac{c}{b}. \tag{36}$$

**Reactive-3 Partner Strategies.** Next, we focus on the case $n=3$. Reactive-3 strategies can be represented as a vector $\mathbf{p} = (p_{CCC}, p_{CCD}, p_{CDC}, p_{CDD}, p_{DCC}, p_{DCD}, p_{DDC}, p_{DDD})$. Again, each entry $p_{\mathbf{h}^{-\mathbf{i}}}$ refers to the player's cooperation probability, depending on the co-player's previous three actions, $\mathbf{h}^{-\mathbf{i}} = (a_{-3}^{-i}, a_{-2}^{-i}, a_{-1}^{-i})$. For the respective partner strategies, we obtain the following characterization.

13

**Theorem 3** (Reactive-3 partner strategies in the donation game)

*A nice reactive-3 strategy* **p** *is a partner strategy if and only if its entries satisfy the conditions*

$$p_{CCC} = 1$$
$$\frac{p_{CDC} + p_{DCD}}{2} \leq 1 - \frac{1}{2} \cdot \frac{c}{b}$$
$$\frac{p_{CCD} + p_{CDC} + p_{DCC}}{3} \leq 1 - \frac{1}{3} \cdot \frac{c}{b}$$
$$\frac{p_{CDD} + p_{DCD} + p_{DDC}}{3} \leq 1 - \frac{2}{3} \cdot \frac{c}{b} \qquad (37)$$
$$\frac{p_{CCD} + p_{CDD} + p_{DCC} + p_{DDC}}{4} \leq 1 - \frac{1}{2} \cdot \frac{c}{b}$$
$$p_{DDD} \leq 1 - \frac{c}{b}$$

As before, the average of certain cooperation probabilities need to be below specific thresholds. However, compared to the case $n=2$, the respective conditions are now somewhat more difficult to interpret. The conditions again become more straightforward if we further restrict attention to reactive-3 counting strategies.

**Corollary 3** (Partners among the reactive-3 counting strategies)

*A nice reactive-3 counting strategy* $\mathbf{r} = (r_3, r_2, r_1, r_0)$ *is a partner strategy if and only if*

$$r_3 = 1 \qquad r_2 \leq 1 - \frac{1}{3} \cdot \frac{c}{b}, \qquad r_1 \leq 1 - \frac{2}{3} \cdot \frac{c}{b}, \qquad r_0 \leq 1 - \frac{c}{b}. \qquad (38)$$

As in the case of $n=2$ we observe here that with each additional defection of the opponent in memory, the focal player reduces its conditional cooperation probability by a constant, in this case $c/(3b)$.

**Partners among the reactive-$n$ counting strategies.** Using the same methods as before, one can in principle also characterize the partners among the reactive-4 or the reactive-5 strategies. However, the respective conditions quickly become unwieldy. In case of the counting strategies, however, the simple pattern in Corollaries 2 and 3 does generalize to arbitrary memory lengths.

**Theorem 4** (Partners among the reactive-$n$ counting strategies)

*A nice reactive-$n$ counting strategy* $\mathbf{r} = (r_k)_{k \in \{n, n-1, \ldots, 0\}}$, *is a partner strategy if and only if*

$$r_n = 1 \qquad and \qquad r_{n-k} \leq 1 - \frac{k}{n} \cdot \frac{c}{b} \quad \text{for } k \in \{1, 2, \ldots, n\}. \qquad (39)$$

### 2.5 Reactive partner strategies in the general prisoner's dilemma

In the previous section, we have characterized the reactive partner strategies for a special case of the prisoner's dilemma, the donation game. In the following, we apply the same methods based on Section 2.3

to analyze the general prisoner's dilemma. For the case of reactive-2 strategies, we obtain the following characterization.

**Theorem 5** (Reactive-2 partner strategies in the prisoner's dilemma)
*A nice reactive-2 strategy* **p** *is a partner strategy if and only if its entries satisfy the conditions*

$$\begin{aligned}
p_{CC} &= 1, \\
(T-P)\,p_{DD} &\leq R-P, \\
(R-S)\,(p_{CD}+p_{DC}) &\leq 3R-2S-T, \\
(T-P)\,p_{DC}+(R-S)\,p_{CD} &\leq 2R-S-P, \\
(T-P)\,(p_{CD}+p_{DC})+(R-S)\,p_{DD} &\leq 3R-S-2P, \\
(T-P)\,p_{CD}+(R-S)\,(p_{CD}+p_{DD}) &\leq 4R-2S-T-P.
\end{aligned} \quad (40)$$

Compared to the donation game, there are now more conditions, and these conditions are somewhat more difficult to interpret. Reassuringly, however, the conditions simplify to the conditions (34) in the special case that the payoff values satisfy $R=b-c$, $S=-c$, $T=b$, and $P=0$. For the case of reactive-3 strategies, the characterization is as follows.

**Theorem 6** (Reactive-3 partner strategies in the prisoner's dilemma)
*A nice reactive-3 strategy* **p** *is a partner strategy if and only if its entries satisfy the conditions in Table 1.*

Given the large number of conditions in Table 1, we do not pursue a similar characterization for $n>3$, even though the same methods remain applicable.

$$
\begin{aligned}
p_{CCC} &= 1, \\
(T-P)(p_{CDD}+p_{DCD}+p_{DDC}) + (R-S)\,p_{DDD} &\le 4R-3P-S \\
(T-P)\,p_{CDC} + (R-S)\,p_{DCD} &\le 2R-P-S \\
(T-P)\,p_{DDD} &\le R-P \\
(T-P)(p_{CCD}+p_{CDD}+p_{DDC}) + (R-S)(p_{CDC}+p_{DCC}+p_{DCD}+p_{DDD}) &\le 8R-3P-4S-T \\
(T-P)\,p_{DCC} + (R-S)(p_{CCD}+p_{CDC}) &\le 3R-P-2S \\
(T-P)(p_{CCD}+p_{DCC}+p_{DDC}) + (R-S)(p_{CDC}+p_{CDD}+p_{DCD}) &\le 6R-3P-3S \\
(T-P)(p_{CCD}+p_{DDC}) + (R-S)(p_{CDC}+p_{CDD}+p_{DCC}+p_{DCD}) &\le 7R-2P-4S-T \\
(T-P)(p_{CCD}+p_{CDD}+p_{DCC}) + (R-S)(p_{DDC}+p_{DDD}) &\le 5R-3P-2S \\
(T-P)(p_{DCD}+p_{DDC}) + (R-S)\,p_{CDD} &\le 3R-2P-S \\
(T-P)\,p_{CCD} + (R-S)(p_{CDD}+p_{DCC}+p_{DDC}) &\le 5R-P-3S-T \\
(T-P)(p_{CCD}+p_{DCC}) + (R-S)(p_{CDD}+p_{DDC}) &\le 4R-2P-2S \\
(T-P)(p_{CDC}+p_{DCD}) + (R-S)(p_{CCD}+p_{CDD}+p_{DCC}+p_{DDD}) &\le 7R-2P-4S-T \\
(T-P)(p_{CDC}+p_{CDD}+p_{DCD}) + (R-S)(p_{CCD}+p_{DCC}+p_{DDC}+p_{DDD}) &\le 8R-3P-4S-T \\
(T-P)(p_{CDC}+p_{DCC}+p_{DCD}) + (R-S)(p_{CCD}+p_{CDD}+p_{DDC}) &\le 6R-3P-3S \\
(T-P)(p_{CCD}+p_{CDD}+p_{DCC}+p_{DDC}) + (R-S)(p_{CDC}+p_{DCD}+p_{DDD}) &\le 7R-4P-3S \\
(R-S)(p_{CCD}+p_{CDC}+p_{DCC}) &\le 4R-3S-T \\
(T-P)(p_{CCD}+p_{CDD}) + (R-S)(p_{DCC}+p_{DDC}+p_{DDD}) &\le 6R-2P-3S-T \\
(T-P)(p_{CDC}+p_{CDD}+p_{DCC}+p_{DCD}) + (R-S)(p_{CCD}+p_{DDC}+p_{DDD}) &\le 7R-4P-3S
\end{aligned}
$$

**Table 1:** Necessary and sufficient conditions for a nice reactive-3 strategy to be a partner in the prisoner's dilemma.

# 3 Appendix: Proofs

## 3.1 Proof of Lemma 1: Akin's lemma

*Proof.* The proof is based on a similar argument as the proof of Eq. (18), showing that different ways of calculating payoffs are equivalent. Let us first introduce some notation. Let $\mathbf{m^1}$ be the memory-$n$ strategy of player 1. For $t \geq n$ and the given strategy of player 2, let $\mathbf{v}(t) = (v_\mathbf{h})_{\mathbf{h} \in H}$ be the probability that player 1 observes the $n$-history $\mathbf{h}$ after players have made their $t$-th decision. By assumption, we can compute the limiting distribution

$$\mathbf{v} = \lim_{\tau \to \infty} \frac{1}{\tau} \sum_{t=n}^{n+\tau-1} \mathbf{v}(t). \tag{41}$$

Moreover, let $\rho^i(t)$ be player i's cooperation probability in round $t$. For $t \geq n+1$, we obtain

$$\rho^1(t) = \langle \mathbf{v}(t-1), \mathbf{m^1} \rangle = \langle \mathbf{v}(t+k-1), \mathbf{m^{k-Rep}} \rangle. \tag{42}$$

That is, we either need to know how likely each $n$-history occurred at time $t-1$, and then we compute how likely player 1 is to cooperate in the next round, based on player 1's strategy. Or, we need to know how likely each $n$-history occurred after round $t+k-1$; and then we compute the correct probability by assuming player 1 cooperates in the next round if and only if the player cooperated $k$ rounds before. Eq. (42) gives us two different ways to compute player 1's average payoff across all rounds,

$$\rho^1 := \lim_{\tau \to \infty} \frac{1}{\tau} \sum_{t=1}^{\tau} \rho^1(t). \tag{43}$$

The first way is to take

$$\begin{aligned}
\rho^1 &= \lim_{\tau \to \infty} \frac{1}{\tau} \sum_{t=1}^{\tau} \rho^i(t) = \lim_{\tau \to \infty} \frac{1}{\tau} \sum_{t=n+1}^{n+\tau} \rho^i(t) \\
&= \lim_{\tau \to \infty} \frac{1}{\tau} \sum_{t=n+1}^{n+\tau} \langle \mathbf{v}(t-1), \mathbf{m^1} \rangle = \langle \lim_{\tau \to \infty} \frac{1}{\tau} \sum_{t=n+1}^{n+\tau} \mathbf{v}(t-1), \mathbf{m^1} \rangle = \langle \mathbf{v}, \mathbf{m^1} \rangle.
\end{aligned}$$

In particular, because $\langle \mathbf{v}, \mathbf{m^1} \rangle$ is well-defined, so is the limiting time average $\rho^1$. The second way is to take

$$\begin{aligned}
\rho^1 &= \lim_{\tau \to \infty} \frac{1}{\tau} \sum_{t=1}^{\tau} \rho^i(t) = \lim_{\tau \to \infty} \frac{1}{\tau} \sum_{t=n+1}^{n+\tau} \rho^i(t) \\
&= \lim_{\tau \to \infty} \frac{1}{\tau} \sum_{t=n+1}^{n+\tau} \langle \mathbf{v}(t+k-1), \mathbf{m^{k-Rep}} \rangle = \langle \lim_{\tau \to \infty} \frac{1}{\tau} \sum_{t=n+1}^{n+\tau} \mathbf{v}(t+k-1), \mathbf{m^{k-Rep}} \rangle = \langle \mathbf{v}, \mathbf{m^{k-Rep}} \rangle.
\end{aligned}$$

We conclude $0 = \rho^1 - \rho^1 = \langle \mathbf{v}, \mathbf{m^1} \rangle - \langle \mathbf{v}, \mathbf{m^{k-Rep}} \rangle = \langle \mathbf{v}, \mathbf{m^1} - \mathbf{m^{k-Rep}} \rangle$. □

## 3.2 Proof of Lemma 2: Sufficiency of testing self-reactive strategies

*Proof.* The proof uses similar arguments as in a study by Park on alternating games *et al* [12]. For the given game between player 1 (with arbitrary strategy $\sigma^1$) and player 2 (with reactive-$n$ strategy $\mathbf{p^2}$), let $v_\mathbf{h}(t)$ denote the probability to observe an $n$-history $\mathbf{h}$ at time $t \geq n$. By assumption, the following time averages are well-defined,

$$v_\mathbf{h} := \lim_{\tau \to \infty} \frac{1}{\tau} \sum_{t=n}^{n+\tau-1} v_\mathbf{h}(t) \tag{44}$$

Moreover, for any $t \geq n$ and $\mathbf{h} \in H$, let $\sigma_\mathbf{h}^1(t)$ denote the conditional probability that player 1 cooperates at time $t+1$, given the $n$-history after round $t$ is $\mathbf{h}$. Depending on $(\sigma_\mathbf{h}^1(t))$ and $\mathbf{v}$, we define an associated self-reactive strategy $\tilde{\mathbf{p}}^1$ for player 1. For any given history $\mathbf{h^1} \in H^1$, the corresponding probability $\tilde{p}_{\mathbf{h^1}}^1$ is defined as an implicit solution of the equation

$$\left( \sum_{\mathbf{h^2} \in H^2} v_{(\mathbf{h^1,h^2})} \right) \tilde{p}_{\mathbf{h^1}}^1 = \sum_{\mathbf{h^2} \in H^2} \left( \lim_{\tau \to \infty} \frac{1}{\tau} \sum_{t=n}^{n+\tau-1} v_{(\mathbf{h^1,h^2})}(t) \cdot \sigma_{(\mathbf{h^1,h^2})}^1(t) \right). \tag{45}$$

Note that for each $\mathbf{h} \in H$, the limit in the bracket on the right hand side exists, for otherwise the limits $v_\mathbf{h}$ according to Eq. (44) would not exist. Also note that if the bracket on the left hand's side is zero, the right hand side must be zero, and $\tilde{p}_{\mathbf{h^1}}^1$ can be chosen arbitrarily. Only if the bracket on the left hand side is positive, $\tilde{p}_{\mathbf{h^1}}^1$ is uniquely defined.

We are going to show: If player 1 uses $\tilde{\mathbf{p}}^1$ instead of $\sigma^1$, then $\mathbf{v}$ defined by Eq. (44) is an invariant distribution of the corresponding transition matrix $M$ defined by Eq. (12) (hence it is also the limiting distribution of the resulting game if the first $n$ moves are chosen accordingly). For simplicity, we show the required relationship $\mathbf{v} = \mathbf{v}M$ for one of the $2^{2^n}$ equations. For the one equation we show, we consider the history according to which everyone fully cooperates, $\mathbf{h_C} = (\mathbf{h_C^1}, \mathbf{h_C^2})$. For an arbitrary $n$-history $\mathbf{h}^i = (a_{-n}^i, \ldots, a_{-i}^i)$, we say the $n$-history $\tilde{\mathbf{h}}^i = (\tilde{a}_{-n}^i, \ldots, \tilde{a}_{-1}^i)$ *is a possible successor of* $\mathbf{h}$ if $\tilde{a}_{-t}^i = a_{-t+1}^i$ for $t \in \{2, \ldots, n\}$. To indicate successorship, we define a function $e_{\mathbf{h},\tilde{\mathbf{h}}}$ that is one if $\tilde{\mathbf{h}}$ is a possible successor of $\mathbf{h}$, and zero otherwise. By definition of $v_\mathbf{h}(t)$, $\sigma_\mathbf{h}^1(t)$, and $p_\mathbf{h}^2(t)$, we obtain for $t \geq n$

$$v_{(\mathbf{h_C^1,h_C^2})}(t+1) = \sum_{\mathbf{h^1} \in H^1} \sum_{\mathbf{h^2} \in H^2} v_{(\mathbf{h^1,h^2})}(t) \cdot \sigma_{(\mathbf{h^1,h^2})}^1(t) \cdot p_{\mathbf{h^1}}^2 \cdot e_{\mathbf{h^1,h_C^1}} \cdot e_{\mathbf{h^2,h_C^2}}. \tag{46}$$

If we sum up this equation from time $t=n$ to $t=n+\tau-1$, divide by $\tau$, and rearrange the terms, we obtain

$$\frac{1}{\tau} \sum_{t=n}^{n+\tau-1} v_{(\mathbf{h_C^1,h_C^2})}(t+1) = \sum_{\mathbf{h^1} \in H^1} \sum_{\mathbf{h^2} \in H^2} \left( \frac{1}{\tau} \sum_{t=n}^{n+\tau-1} v_{(\mathbf{h^1,h^2})}(t) \cdot \sigma_{(\mathbf{h^1,h^2})}^1(t) \right) \cdot p_{\mathbf{h^1}}^2 \cdot e_{\mathbf{h^1,h_C^1}} \cdot e_{\mathbf{h^2,h_C^2}}. \tag{47}$$

Taking the limit $\tau \to \infty$, and taking into account the relationships (44) and (45), this simplifies to

$$v_{(\mathbf{h_C^1,h_C^2})} = \sum_{\mathbf{h^1} \in H^1} \sum_{\mathbf{h^2} \in H^2} v_{(\mathbf{h^1,h^2})} \cdot \left( \tilde{p}_{\mathbf{h^1}}^1 e_{\mathbf{h^1,h_C^1}} \right) \cdot \left( p_{\mathbf{h^1}}^2 e_{\mathbf{h^2,h_C^2}} \right). \tag{48}$$

18

By using the definition of transition probabilities in (12), this expression further simplifies to

$$v_{\mathbf{h_C}} = \sum_{\mathbf{h}} v_{\mathbf{h}} \cdot M_{\mathbf{h},\mathbf{h_C}} \tag{49}$$

That is, out of the $2^{2n}$ individual equations in the linear system $\mathbf{v} = \mathbf{v}M$, we have verified the equation for the probability to observe full cooperation $\mathbf{h_C}$ after one round. All other equations follow analogously. □

### 3.3 Proof of Theorem 1: Sufficiency of pure self-reactive strategies

By Lemma 2, there exists a best response to $\mathbf{p}$ within the self-reactive $n$ strategies. It remains to show that this best response $\tilde{\mathbf{p}}$ can be chosen to be pure. The proof follows from a series of auxiliary results. The first such result uses an insight by Press & Dyson [9]. They showed that given the transition matrix of a game among two memory-1 players, one can compute the players' payoffs by considering determinants of certain associated matrices. Herein, we apply their method to the transition matrix $\tilde{M} = (\tilde{M}_{\mathbf{h},\mathbf{h}'})$ according to Eq. (26) for a given self-reactive strategy $\tilde{\mathbf{p}} \in \mathcal{S}_n$. For some fixed $n$-history $\mathbf{h}'$, we define an associated matrix $\tilde{M}_{\mathbf{h}'}$ that one obtains from $\tilde{M}$ with the following two steps:

1. Subtract the $2^n \times 2^n$ identity matrix $I$ from $\tilde{M}$.

2. In the resulting matrix, replace the last column by a column that only contains zeros, except for the row corresponding to the history $\mathbf{h}'$, for which the entry is one.

These matrices $\tilde{M}_{\mathbf{h}'}$ can be used to compute the invariant distribution of the original matrix $\tilde{M}$ as follows.

**Auxiliary result 1:** Let $\tilde{\mathbf{p}} \in \mathcal{S}_n$ be such that its transition matrix $\tilde{M}$ according to Eq. (26) has a unique invariant distribution $\tilde{\mathbf{v}} = (\tilde{v}_{\mathbf{h}^1})_{\mathbf{h}^1 \in H^1}$. Then for all $\mathbf{h}' \in H^1$ we have

$$\tilde{v}_{\mathbf{h}'} = \frac{\det(\tilde{M}_{\mathbf{h}'})}{\sum_{\mathbf{h}^1 \in H^1} \det(\tilde{M}_{\mathbf{h}^1})}. \tag{50}$$

*Proof of Auxiliary result 1.* The result follows from Press & Dyson's formula for the dot product of the invariant distribution $\tilde{\mathbf{v}}$ with an arbitrary vector $\mathbf{f}$, by taking the vector $\mathbf{f}$ to be the unit vector with only the entry for history $\mathbf{h}'$ being one. □

Based on this first auxiliary result, we have an explicit representation of the payoff function $\pi^1(\tilde{\mathbf{p}}, \mathbf{p})$ that describes the payoff of a self-reactive player with strategy $\tilde{\mathbf{p}}$ against a reactive player with strategy $\mathbf{p}$. Specifically, by plugging Eq. (50) into (27), we obtain

$$\pi^1(\tilde{\mathbf{p}}, \mathbf{p}) = \frac{\sum_{\mathbf{h}^1 \in H^1} \det(\tilde{M}_{\mathbf{h}^1}) \Big( \tilde{p}_{\mathbf{h}^1} p_{\mathbf{h}^1} \cdot R + \tilde{p}_{\mathbf{h}^1}(1-p_{\mathbf{h}^1}) \cdot S + (1-\tilde{p}_{\mathbf{h}^1}) p_{\mathbf{h}^1} \cdot T + (1-\tilde{p}_{\mathbf{h}^1})(1-p_{\mathbf{h}^1}) \cdot P \Big)}{\sum_{\mathbf{h}^1 \in H^1} \det(\tilde{M}_{\mathbf{h}^1})}. \tag{51}$$

For our purposes, the following properties of this payoff function will be important.

19

**Auxiliary Result 2:** On its domain, the payoff function $\pi^1(\tilde{\mathbf{p}}, \mathbf{p})$ is a bounded rational function, and both its numerator and denominator are linear in each entry $\tilde{p}_{\mathbf{h}^i}$, for all $\mathbf{h}^i \in H^i$.

*Proof of Auxiliary Result 2.* By its definition, each $\det(\tilde{M}_{\mathbf{h}^i})$ is a polynomial. Moreover, because for each history $\mathbf{h}'$, the cooperation probability $\tilde{p}_{\mathbf{h}'}$ only appears in a single row of $\tilde{M}_{\mathbf{h}^i}$ (and there it appears linearly), it also appears linearly in $\det(\tilde{M}_{\mathbf{h}^i})$. Finally, we note that $\det(\tilde{M}_{\mathbf{h}^i})$ does not depend on $\tilde{p}_{\mathbf{h}^i}$. To see this, we can compute $\det(\tilde{M}_{\mathbf{h}^i})$ using Laplace expansion along the last column. As a result, we obtain that this determinant is up to its sign equal to the determinant of the matrix one obtains from $\tilde{M}_{\mathbf{h}^i}$ by deleting the last column, and the row $\mathbf{h}^i$ (which is the only row of $\tilde{M}_{\mathbf{h}^i}$ that contains $\tilde{p}_{\mathbf{h}^i}$).

Finally, we note that the payoff function is bounded, because as an average payoff per round, payoffs need to be between $T$ and $S$. Taken together, these observations imply the result for $\pi^1(\tilde{\mathbf{p}}, \mathbf{p})$. □

The following result describes a useful property of bounded linear rational functions.

**Auxiliary Result 3:** Suppose $g, h : [0,1]^k \to \mathbb{R}$ and suppose both $g(\mathbf{x})$ and $h(\mathbf{x})$ are linear in each component of $\mathbf{x} = (x_1, \ldots, x_k)$. Moreover, suppose $f := g/h$ is bounded on $[0,1]^k$. For a given $\mathbf{x}$ and $j \in \{1, \ldots, k\}$, we define an associated function $f_{\mathbf{x},j} : [-x_j, 1-x_j] \to \mathbb{R}$ by only varying the $j$-th component, $f_{\mathbf{x},j}(t) = f(x_1, \ldots, x_j + t, \ldots, x_k)$. Then for all $\mathbf{x} \in [0,1]^k$ and $j$, the function $f_{\mathbf{x},j}(t)$ is either monotonically increasing, monotonically decreasing, or constant.

*Proof of Auxiliary Result 3.* Let $g(\mathbf{x}) := a_0 + a_1 x_1 + \ldots + a_k x_k$ and $h(\mathbf{x}) := b_0 + b_1 x_1 + \ldots + b_k x_k$, and consider some arbitrary but fixed $\mathbf{x} \in [0,1]^k$ and $j$. We compute

$$f'_{\mathbf{x},j}(t) = \frac{\partial}{\partial t} f(x_1, \ldots, x_j + t, \ldots, x_k) = \frac{a_j \left(\sum_{i \neq j} b_i x_i\right) - b_j \left(\sum_{i \neq j} a_i x_i\right)}{\left(b_0 + b_1 x_1 + \ldots + b_j(x_j + t) + \ldots + b_k x_k\right)^2}. \tag{52}$$

Because $f$ is bounded on the entire domain, this expression for $f'_{\mathbf{x},j}(t)$ is finite. It hence follows fthat the denominator of $f'_{\mathbf{x},j}(t)$ is always positive, and that the numerator is independent of $t$. Thus, depending on the sign of the numerator, $f'_{\mathbf{x},j}(t)$ is either monotonically increasing, monotonically decreasing, or constant. □

After these preparations, we are ready to prove the main result.

*Proof of Theorem 1.* For a given reactive strategy $\mathbf{p} \in \mathcal{R}_n$, let the self-reactive $\tilde{\mathbf{p}} \in \mathcal{S}_n$ be a best response. Suppose there is some history $\mathbf{h}'$ such that $0 < \tilde{p}_{\mathbf{h}} < 1$. It follows from the Auxiliary Results 2 and 3 that $\pi^1(\tilde{\mathbf{p}}, \mathbf{p})$ is either monotonically increasing, monotonically decreasing, or constant in $\tilde{p}_{\mathbf{h}'}$. If it was increasing or decreasing, we end up with a contradiction, because no local improvement should be possible for a best response. Therefore, $\pi^1(\tilde{\mathbf{p}}, \mathbf{p})$ must be independent of $\tilde{p}_{\mathbf{h}'}$, and hence we can set $\tilde{p}_{\mathbf{h}'} = 0$ or $\tilde{p}_{\mathbf{h}'} = 1$ without changing $\pi^1(\tilde{\mathbf{p}}, \mathbf{p})$. By iteratively applying this reasoning to all histories $\mathbf{h}$ for which $0 < \tilde{p}_{\mathbf{h}} < 1$, we obtain the desired result. □

### 3.4 Proof of Theorem 2: Reactive-2 partner strategies in the donation game

*Proof.* Given that player 1 uses a nice reactive-2 strategy $\mathbf{p} = (1, p_{CD}, p_{DC}, p_{DD})$, the claim is true if and only if it is true for all deviation towards the sixteen pure self-reactive-2 strategies $\tilde{\mathbf{p}} \in \{0,1\}^{16}$. In the following, we enumerate these sixteen strategies, $\{\tilde{\mathbf{p}}_0, \ldots, \tilde{\mathbf{p}}_{15}\}$, by interpreting them as binary numbers,

$$\tilde{\mathbf{p}} = (\tilde{p}_{CC}, \tilde{p}_{CD}, \tilde{p}_{DC}, \tilde{p}_{DD}) \mapsto \tilde{p}_{CC} \cdot 2^3 + \tilde{p}_{CD} \cdot 2^2 + \tilde{p}_{DC} \cdot 2^1 + p_{DD} \cdot 2^0. \qquad (53)$$

In particular, $\texttt{ALLD} = (0,0,0,0)$ is mapped to the number $j=0$, and $\texttt{ALLC} = (1,1,1,1)$ is mapped to $j=15$. The possible payoffs against the reactive strategy $\mathbf{p}$ can be computed by Eq. (27), which yields

$$\pi^1(\tilde{\mathbf{p}}_\mathbf{j}, \mathbf{p}) = p_{DD} \cdot b \qquad \text{for } j \in \{0, 2, 4, 6, 8, 10, 12, 14\}$$

$$\pi^1(\tilde{\mathbf{p}}_\mathbf{j}, \mathbf{p}) = \frac{p_{CD} + p_{DC} + p_{DD}}{3} \cdot b - \frac{1}{3} \cdot c \qquad \text{for } j \in \{1, 9\}$$

$$\pi^1(\tilde{\mathbf{p}}_\mathbf{j}, \mathbf{p}) = \frac{1 + p_{CD} + p_{DC} + p_{DD}}{4} \cdot b - \frac{1}{2} \cdot c \qquad \text{for } j \in \{3\}$$

$$\pi^1(\tilde{\mathbf{p}}_\mathbf{j}, \mathbf{p}) = \frac{p_{CD} + p_{DC}}{2} \cdot b - \frac{1}{2} \cdot c \qquad \text{for } j \in \{4, 5, 12, 13\}$$

$$\pi^1(\tilde{\mathbf{p}}_\mathbf{j}, \mathbf{p}) = \frac{1 + p_{CD} + p_{DC}}{3} \cdot b - \frac{2}{3} \cdot c \qquad \text{for } j \in \{6, 7\}$$

$$\pi^1(\tilde{\mathbf{p}}_\mathbf{j}, \mathbf{p}) = b - c \qquad \text{for } j \in \{8, 9, 10, 11, 12, 13, 14, 15\}$$

In this list, some strategy indices $j$ appear multiple times. Those instances correspond to strategies that have multiple invariant distributions (such as the strategy 1-round repeat, with $j=10$). For those strategies, we have computed the payoffs for all possible initial $n$-histories. Requiring the payoffs in this list to be at most the mutual cooperation payoff $b-c$, we get the following unique conditions,

$$p_{DD} \leq 1 - \frac{c}{b}, \qquad \frac{p_{CD} + p_{DC}}{2} \leq 1 - \frac{1}{2}\frac{c}{b}, \qquad \frac{p_{CD} + p_{DC} + p_{DD}}{3} \leq 1 - \frac{2}{3}\frac{c}{b}.$$

Because the last condition is implied by the first two, we end up with the conditions in (34). □

## 3.5 Proof of Theorem 3: Reactive-3 partner strategies in the donation game

*Proof.* The proof is similar to the previous one. Again, enumerating the 256 pure self-reactive 3 strategies $\tilde{\mathbf{p}}$ by interpreting the strategy as a binary number, we obtain the following payoffs.

$\pi^1(\tilde{\mathbf{p}}_j, \mathbf{p}) = b\, p_{DDD}$ for $j \in \{0, 2, 4, 6, \ldots, 250, 252, 254\}$

$\pi^1(\tilde{\mathbf{p}}_j, \mathbf{p}) = \frac{p_{CDD}+p_{DCD}+p_{DDC}+p_{DDD}}{4} b - \frac{1}{4} c$ for $j \in \{1, 9, 33, 41, 65, 73, 97, 105, 129, 137, 161, 169, 193, 201, 225, 233\}$

$\pi^1(\tilde{\mathbf{p}}_j, \mathbf{p}) = \frac{p_{CCD}+p_{CDD}+p_{DCC}+p_{DDC}+p_{DDD}}{5} b - \frac{2}{5} c$ for $j \in \{3, 7, 35, 39, 131, 135, 163, 167\}$

$\pi^1(\tilde{\mathbf{p}}_j, \mathbf{p}) = \frac{p_{CDC}+p_{DCD}}{2} b - \frac{1}{2} c$ for $j \in \{4-7, 12-15, 20-23, 28-31, 68-71, 76-79, 84-87, 92-95, 132-135, 140-143, 148-151, 156-159, 196-199, 204-207, 212-215, 220-223\}$

$\pi^1(\tilde{\mathbf{p}}_j, \mathbf{p}) = \frac{1+p_{CCD}+p_{CDD}+p_{DCC}+p_{DDC}+p_{DDD}}{6} b - \frac{1}{2} c$ for $j \in \{11, 15, 43, 47\}$

$\pi^1(\tilde{\mathbf{p}}_j, \mathbf{p}) = \frac{p_{CDD}+p_{DCD}+p_{DDC}}{3} b - \frac{1}{3} c$ for $j \in \{16, 17, 24, 25, 48, 49, 56, 57, 80, 81, 88, 89, 112, 113, 120, 121, 144, 145, 152, 153, 176, 177, 184, 185, 208, 209, 216, 217, 240, 241, 248, 249\}$

$\pi^1(\tilde{\mathbf{p}}_j, \mathbf{p}) = \frac{p_{CCD}+p_{CDD}+p_{DCC}+p_{DDC}}{4} b - \frac{1}{2} c$ for $j \in \{18, 19, 22, 23, 50, 51, 54, 55, 146, 147, 150, 151, 178, 179, 182, 183\}$

$\pi^1(\tilde{\mathbf{p}}_j, \mathbf{p}) = \frac{1+p_{CCD}+p_{CDD}+p_{DCC}+p_{DDC}}{5} b - \frac{3}{5} c$ for $j \in \{26, 27, 30, 31, 58, 59, 62, 63\}$

$\pi^1(\tilde{\mathbf{p}}_j, \mathbf{p}) = \frac{p_{CCD}+p_{CDC}+p_{CDD}+p_{DCC}+p_{DCD}+p_{DDC}+p_{DDD}}{7} b - \frac{3}{7} c$ for $j \in \{37, 67, 165, 195\}$

$\pi^1(\tilde{\mathbf{p}}_j, \mathbf{p}) = \frac{1+p_{CCD}+p_{CDC}+p_{CDD}+p_{DCC}+p_{DCD}+p_{DDC}+p_{DDD}}{8} b - \frac{1}{2} c$ for $j \in \{45, 75\}$

$\pi^1(\tilde{\mathbf{p}}_j, \mathbf{p}) = \frac{p_{CCD}+p_{CDC}+p_{CDD}+p_{DCC}+p_{DCD}+p_{DDC}}{6} b - \frac{1}{2} c$ for $j \in \{52, 53, 82, 83, 180, 181, 210, 211\}$

$\pi^1(\tilde{\mathbf{p}}_j, \mathbf{p}) = \frac{1+p_{CCD}+p_{CDC}+p_{CDD}+p_{DCC}+p_{DCD}+p_{DDC}}{7} b - \frac{4}{7} c$ for $j \in \{60, 61, 90, 91\}$

$\pi^1(\tilde{\mathbf{p}}_j, \mathbf{p}) = \frac{p_{CCD}+p_{CDC}+p_{DCC}}{3} b - \frac{2}{3} c$ for $j \in \{96-103, 112-119, 224-231, 240-247\}$

$\pi^1(\tilde{\mathbf{p}}_j, \mathbf{p}) = \frac{1+p_{CCD}+p_{CDC}+p_{DCC}}{4} b - \frac{3}{4} c$ for $j \in \{104-111, 120-127\}$

$\pi^1(\tilde{\mathbf{p}}_j, \mathbf{p}) = b - c$ for $j \in \{128, 129, 130, \ldots, 255\}$

Requiring these payoffs to be at most equal to the mutual cooperation payoff $b-c$ gives

$$p_{DDD} \leq 1 - \frac{c}{b}, \qquad \frac{p_{CDC}+p_{DCD}}{2} \leq 1 - \frac{1}{2} \cdot \frac{c}{b}, \qquad \frac{p_{CDD}+p_{DCD}+p_{DDC}}{3} \leq 1 - \frac{2}{3} \cdot \frac{c}{b},$$

$$\frac{p_{CCD}+p_{CDC}+p_{DCC}}{3} \leq 1 - \frac{1}{3} \cdot \frac{c}{b}, \qquad \frac{p_{CCD}+p_{CDD}+p_{DCC}+p_{DDC}}{4} \leq 1 - \frac{1}{2} \cdot \frac{c}{b},$$

$$\frac{p_{CDD}+p_{DCD}+p_{DDC}+p_{DDD}}{4} \leq 1 - \frac{3}{4} \cdot \frac{c}{b}, \qquad \frac{p_{CCD}+p_{CDC}+p_{CDD}+p_{DCC}+p_{DCD}+p_{DDC}+p_{DDD}}{7} \leq 1 - \frac{4}{7} \cdot \frac{c}{b},$$

$$\frac{p_{CCD}+p_{CDD}+p_{DCC}+p_{DDC}+p_{DDD}}{5} \leq 1 - \frac{3}{5} \cdot \frac{c}{b}, \qquad \frac{p_{CCD}+p_{CDC}+p_{CDD}+p_{DCC}+p_{DCD}+p_{DDC}}{6} \leq 1 - \frac{1}{2} \cdot \frac{c}{b}.$$

The statement follows by noting that the five conditions in the first two rows imply the four other conditions.

## 3.6 Proof of Theorem 4: Reactive-$n$ counting strategies in the donation game

Before we go into the details of the proof, we first start with two useful observations.

1. Assume player 1 adopts a given self-reactive strategy $\tilde{\mathbf{p}}$ and player 2 adopts the reactive-$n$ strategy $\mathbf{r} = (r_k)_{k \in \{n,\ldots,0\}}$. For the resulting game, suppose $\mathbf{v}$ is the limiting distribution according to Eq. (14). Then it is useful to express $\mathbf{v}$ in terms of what the counting player can remember. To this end, let $H_k^1$ be the set of $n$-histories according to which player 1 has cooperated exactly $k$ times,

$$H_k^1 = \left\{ \mathbf{h^1} \in H^1 \,\middle|\, |\mathbf{h^1}| = k \right\}. \tag{54}$$

Accordingly, let $\mathbf{u} = (u_k)_{k \in \{0,\ldots,n\}}$ be the distribution that summarizes how often, on average, player 1 cooperates $j$ times during $n$ consecutive rounds,

$$u_k^1 = \sum_{\mathbf{h^1} \in H_k^1} v_{\mathbf{h^1}}. \tag{55}$$

In particular, the entries of $\mathbf{u}$ are normalized,

$$\sum_{k=0}^{n} u_k^1 = 1. \tag{56}$$

Moreover, the average cooperation rate of the two players can be written as

$$\rho^1 = \sum_{k=0}^{n} \frac{k}{n} u_k^1 \quad \text{and} \quad \rho^2 = \sum_{k=0}^{n} r_k \, u_k^1. \tag{57}$$

Because payoffs in the donation game only depend on the players' average cooperation rates (but not on the timing of cooperation), we conclude that player 1's payoff is

$$\pi^1(\tilde{\mathbf{p}}, \mathbf{r}) = \sum_{k=0}^{n} (r_k \, b - \frac{k}{n} c) \, u_k^1. \tag{58}$$

2. There is a set of strategies for which payoffs are particularly easy to compute. We refer to them as simple periodic strategies, $\sigma_k$ with $k \in \{0,\ldots,n\}$. A player with strategy $\sigma_k$ cooperates in round $t$ if and only if

$$t - 1 \mod n < k. \tag{59}$$

That is, such a player cooperates in the first $k$ rounds, then defects for $n-k$ rounds, then cooperates for another $k$ rounds, only to defect in the $n-k$ subsequent rounds, etc. Such strategies are interesting for two reasons. First, they all can be interpreted as a round-$n$ repeat strategy $\tilde{\mathbf{p}}^{\mathbf{n}-\mathrm{Rep}}$, as defined by (11). During the initial $n$ rounds, they cooperate according to Eq. (59); thereafter, they simply

repeat whatever they have done $n$ rounds ago. Second, players with strategy $\sigma_k$ always act in such a way that according to any resulting $n$-history, they have cooperated exactly $k$ times during the last $n$ rounds. As a result, if player 1 adopts such a strategy in a donation game against a player with a reactive-$n$ counting strategy $\mathbf{r}$, then player 1's average payoff is

$$\pi^1(\sigma_k, \mathbf{r}) = r_k\, b - \frac{k}{n}\, c. \tag{60}$$

After these observations, we are ready for the actual proof.

*Proof of Theorem 4.*

($\Rightarrow$) Suppose the reactive-$n$ counting strategy $\mathbf{r}$ is a partner. Because it is nice, it cooperates against an unconditional cooperator, and hence $r_n = 1$. Because it is a Nash equilibrium, player 1 must not have an incentive to deviate towards any of the simple periodic strategies $\sigma_k$. By Eq. (60), this means that for all $k \in \{0, \ldots, n\}$ we have

$$r_k\, b - \frac{k}{n}\, c \leq b - c. \tag{61}$$

These conditions are equivalent to $r_{n-k} \leq 1 - \frac{k}{n}\frac{c}{b}$, the inequalities in (39).

($\Leftarrow$) Because $\mathbf{r}$ is nice, $r_n = 1$. The proof is now by contradiction; suppose the conditions in (39) hold, but $\mathbf{r}$ is not a Nash equilibrium. Then there needs to be some self-reactive $\tilde{\mathbf{p}}$ such that $\pi^1(\tilde{\mathbf{p}}, \mathbf{r}) > b - c$. It follows that

$$\begin{aligned}
0 &< \pi^1(\tilde{\mathbf{p}}, \mathbf{r}) - (b-c) \\
&\stackrel{(56),(58)}{=} \sum_{k=0}^{n}\left(r_k\, b - \frac{k}{n}\, c\right) u^1_k - \sum_{k=0}^{n}(b-c) u^1_k \\
&= (r_n - 1) b\, u_n + \sum_{k=0}^{n-1}\left((r_k - 1) b + \frac{n-k}{n} c\right) u^1_k \\
&= b \cdot \sum_{k=1}^{n}\underbrace{\left(r_{n-k} - \left(1 - \frac{k}{n}\frac{c}{b}\right)\right)}_{\leq 0 \text{ by Eq. (39)}} u^1_{n-k} \leq 0.
\end{aligned} \tag{62}$$

We end up with $0 < 0$, a contradiction.

$\square$

## 3.7 Proof of Theorem 5: Reactive-2 partner strategies in the prisoner's dilemma

*Proof.* The proof is analogous to the proof of Theorem 2 for the donation game. For the general prisoner's dilemma, the payoffs of the 16 pure self-reactive-2 strategies are

$$\pi^1(\tilde{\mathbf{p}}_\mathbf{j}, \mathbf{p}) = P(1-p_{DD}) + Tp_{DD} \qquad \text{for } i \in \{0, 2, 4, 6, 8, 10, 12, 14\}$$

$$\pi^1(\tilde{\mathbf{p}}_\mathbf{j}, \mathbf{p}) = \frac{Rp_{DD} + S(1 - p_{DD}) + T(p_{CD} + p_{DC}) + P(2 - p_{CD} - p_{DC})}{3} \qquad \text{for } i \in \{1, 9\}$$

$$\pi^1(\tilde{\mathbf{p}}_\mathbf{j}, \mathbf{p}) = \frac{R(p_{DC} + p_{DD}) + S(2 - p_{DC} - p_{DD}) + T(p_{CD} + 1) + P(1 - p_{CD})}{4} \qquad \text{for } i \in \{3\}$$

$$\pi^1(\tilde{\mathbf{p}}_\mathbf{j}, \mathbf{p}) = \frac{Rp_{CD} + S(1 - p_{CD}) + Tp_{DC} + P(1 - p_{DC})}{2} \qquad \text{for } i \in \{4, 5, 12, 13\}$$

$$\pi^1(\tilde{\mathbf{p}}_\mathbf{j}, \mathbf{p}) = \frac{R(p_{CD} + p_{DC}) + S(2 - p_{CD} - p_{DC}) + T}{3} \qquad \text{for } i \in \{6, 7\}$$

$$\pi^1(\tilde{\mathbf{p}}_\mathbf{j}, \mathbf{p}) = R \qquad \text{for } i \in \{8, 9, 10, 11, 12, 13, 14, 15\}$$

By requiring these expressions to be at most equal to $R$, we obtain

$$\begin{aligned}
(T - P)\, p_{DD} &\leq R - P, \\
(R - S)\,(p_{CD} + p_{DC}) &\leq 3R - 2S - T, \\
(T - P)\, p_{DC} + (R - S)\, p_{CD} &\leq 2R - S - P, \\
(T - P)\,(p_{CD} + p_{DC}) + (R - S)\, p_{DD} &\leq 3R - S - 2P, \\
(T - P)\, p_{CD} + (R - S)\,(p_{CD} + p_{DD}) &\leq 4R - 2S - P - T.
\end{aligned}$$

□

## 3.8 Proof of Theorem 6: Reactive-3 partner strategies in the prisoner's dilemma

Again, we compute payoffs for all 256 self-reactive-3 strategies. The expressions are given below,

$$\pi^1(\tilde{\mathbf{p}}_j, \mathbf{p}) = \frac{(T-P)(p_{CDD} + p_{DCD} + p_{DDC}) + 3P + (R-S)p_{DDD} + S}{4} \quad \text{for } j \in \{1, 9, 33, 41, 65, 73, 97, 105, 129, 137, 161, 169, 193, 201, 225, 233\}$$

$$\pi^1(\tilde{\mathbf{p}}_j, \mathbf{p}) = \frac{(T-P)p_{CDC} + P + (R-S)p_{DCD} + S}{2} \quad \text{for } j \in \{4-7, 12-15, 20-23, 28-31, 68-71, 76-79, 84-87, 92-95, 132-135, 140-143, 148-151, 156-159, 196-199, 204-207, 212-215, 220-223\}$$

$$\pi^1(\tilde{\mathbf{p}}_j, \mathbf{p}) = -P(p_{DDD} - 1) + Tp_{DDD} \quad \text{for } j \in \{0, 2, 4, \ldots, 252, 254\}$$

$$\pi^1(\tilde{\mathbf{p}}_j, \mathbf{p}) = \frac{(T-P)(p_{CCD} + p_{CDD} + p_{DDC}) + 3P + (R-S)(p_{CDC} + p_{DCC} + p_{DCD} + p_{DDD}) + 4S + T}{8} \quad \text{for } j \in \{45\}$$

$$\pi^1(\tilde{\mathbf{p}}_j, \mathbf{p}) = \frac{(T-P)p_{DCC} + P + (R-S)(p_{CDC} + p_{CCD}) + 2S}{3} \quad \text{for } j \in \{96-103, 112-119, 224-231, 240-247\}$$

$$\pi^1(\tilde{\mathbf{p}}_j, \mathbf{p}) = \frac{(T-P)(p_{CCD} + p_{DCC} + p_{DDC}) + 3P + (R-S)(p_{CDC} + p_{CDD} + p_{DCD}) + 3S}{6} \quad \text{for } j \in \{52, 53, 180, 181\}$$

$$\pi^1(\tilde{\mathbf{p}}_j, \mathbf{p}) = \frac{(T-P)(p_{CCD} + p_{DDC}) + 2P + T + (R-S)(p_{CDC} + p_{CDD} + p_{DCC} + p_{DCD}) + 4S}{7} \quad \text{for } j \in \{60, 61\}$$

$$\pi^1(\tilde{\mathbf{p}}_j, \mathbf{p}) = \frac{(T-P)(p_{CCD} + p_{CDD} + p_{DCC}) + 3P + (R-S)(p_{DDC} + p_{DDD}) + 2S}{5} \quad \text{for } j \in \{3, 7, 35, 39, 131, 135, 163, 167\}$$

$$\pi^1(\tilde{\mathbf{p}}_j, \mathbf{p}) = \frac{(T-P)(p_{DCD} + p_{DDC}) + 2P + (R-S)p_{CDD} + S}{3} \quad \text{for } j \in \{16, 17, 24, 25, 48, 49, 56, 57, 80, 81, 88, 89, 112, 113, 120, 121, 144, 145, 152, 153, 176, 177, 184, 185, 208, 209, 216, 217, 240, 241, 248, 249\}$$

$$\pi^1(\tilde{\mathbf{p}}_j, \mathbf{p}) = R \quad \text{for } j \in \{128, 129, \ldots, 255\}$$

$$\pi^1(\tilde{\mathbf{p}}_j, \mathbf{p}) = \frac{(T-P)p_{CCD} + P + T + (R-S)(p_{CDD} + p_{DCC} + p_{DDC}) + 3S}{5} \quad \text{for } j \in \{26, 27, 30, 31, 58, 59, 62, 63\}$$

$$\pi^1(\tilde{\mathbf{p}}_j, \mathbf{p}) = \frac{(T-P)(p_{CCD} + p_{DCC}) + 2P + (R-S)(p_{CDD} + p_{DDC}) + 2S}{4} \quad \text{for } j \in \{18, 19, 22, 23, 50, 51, 54, 55, 146, 147, 150, 151, 178, 179, 182, 183\}$$

$$\pi^1(\tilde{\mathbf{p}}_j, \mathbf{p}) = \frac{(T-P)(p_{CDC} + p_{DCD}) + 2P + T + (R-S)(p_{CCD} + p_{CDD} + p_{DCC} + p_{DDC}) + 4S}{7} \quad \text{for } j \in \{90, 91\}$$

$$\pi^1(\tilde{\mathbf{p}}_j, \mathbf{p}) = \frac{(T-P)(p_{CDC} + p_{CDD} + p_{DCD}) + 3P + T + (R-S)(p_{CCD} + p_{DCC} + p_{DDC} + p_{DDD}) + 4S}{8} \quad \text{for } j \in \{75\}$$

$$\pi^1(\tilde{\mathbf{p}}_j, \mathbf{p}) = \frac{(T-P)(p_{CDC} + p_{DCC} + p_{DCD}) + 3P + (R-S)(p_{CCD} + p_{CDD} + p_{DDC}) + 3S}{6} \quad \text{for } j \in \{82, 83, 210, 211\}$$

$$\pi^1(\tilde{\mathbf{p}}_j, \mathbf{p}) = \frac{(T-P)(p_{CCD} + p_{CDD} + p_{DCC} + p_{DDC}) + 4P + (R-S)(p_{CDC} + p_{DCD} + p_{DDD}) + 3S}{7} \quad \text{for } j \in \{37, 165\}$$

$$\pi^1(\tilde{\mathbf{p}}_j, \mathbf{p}) = \frac{T + (R-S)(p_{CCD} + p_{CDC} + p_{DCC}) + 3S}{4} \quad \text{for } j \in \{104-111, 120-127\}$$

$$\pi^1(\tilde{\mathbf{p}}_j, \mathbf{p}) = \frac{(T-P)(p_{CCD} + p_{CDD}) + 2P + T + (R-S)(p_{DCC} + p_{DDC} + p_{DDD}) + 3S}{6} \quad \text{for } j \in \{11, 15, 43, 47\}$$

$$\pi^1(\tilde{\mathbf{p}}_j, \mathbf{p}) = \frac{(T-p)(p_{CDC} + p_{CDD} + p_{DCC} + p_{DCD}) + 4P + (R-S)(p_{CCD} + p_{DDC} + p_{DDD}) + 3S}{7} \quad \text{for } j \in \{67, 195\}$$

By requiring the above expressions to be smaller than or equal to $R$, we obtain the inequalities in Table 1.

## 4 Supplementary References



\end{document}